\documentclass[fleqn,usenatbib]{mnras}
\usepackage{natbib}

\usepackage[utf8]{inputenc} 
\usepackage[T1]{fontenc}    
\usepackage{hyperref}       
\usepackage{url}            

\usepackage{natbib}

\usepackage{graphicx}	
\usepackage{amsmath}	
\usepackage{amssymb}	
\usepackage{bm}
\usepackage{hyperref}
\usepackage{breakcites}
\usepackage{url} 
\usepackage{float}
\usepackage{subfigure}
\usepackage{color}
\usepackage{lineno} 
\usepackage{enumitem}  
\usepackage{chngcntr}
\usepackage{cancel}
\usepackage{threeparttable}
\usepackage{tikz}
\usepackage{academicons}
\usepackage{newtxtext,newtxmath}

\newcommand{\eqb}{\begin{eqnarray}}
\newcommand{\eqe}{\end{eqnarray}}
\newcommand{\bi}{\begin{itemize}}
\newcommand{\ei}{\end{itemize}}

\newcommand{\swift}{\emph{Swift}}
\definecolor{frenchblue}{rgb}{0.0, 0.45, 0.73}
\definecolor{burgundy}{rgb}{0.5, 0.0, 0.13}
\definecolor{royalblue}{RGB}{65,105,225}
\definecolor{darkspringgreen}{rgb}{0.09, 0.45, 0.27}

\definecolor{lime}{HTML}{A6CE39}
\DeclareRobustCommand{\orcidicon}{%
	\begin{tikzpicture}
	\draw[lime, fill=lime] (0,0) 
	circle [radius=0.16] 
	node[white] {{\fontfamily{qag}\selectfont \tiny ID}};
	\draw[white, fill=white] (-0.0625,0.095) 
	circle [radius=0.007];
	\end{tikzpicture}
	\hspace{-2mm}
}

\foreach \x in {A, ..., Z}{%
	\expandafter\xdef\csname orcid\x\endcsname{\noexpand\href{https://orcid.org/\csname orcidauthor\x\endcsname}{\noexpand\orcidicon}}
}

\newcommand{\orcid}[1]{\href{https://orcid.org/#1}{\textcolor[HTML]{A6CE39}{\orcidicon}}}
\hypersetup{colorlinks,linkcolor={frenchblue},citecolor={frenchblue},urlcolor={frenchblue}}  
\title[A GRB proton-synchrotron model]{A marginally fast-cooling proton-synchrotron model for prompt GRBs}

\author[Florou et al.]{
Ioulia Florou$^{1}$\thanks{E-mail: iflorou@phys.uoa.gr},
Maria Petropoulou\orcid{0000-0001-6640-0179}$^{1}$\thanks{E-mail: mpetropo@phys.uoa.gr},
Apostolos Mastichiadis\orcid{0000-0001-5217-4801}$^{1}$
\\
$^{1}$Department of Physics, National and Kapodistrian University of Athens, University Campus Zografos, GR 15783, Greece\\
}

\date{Accepted XXX. Received YYY; in original form ZZZ}

\pubyear{2021}

\begin{document}

\label{firstpage}
\pagerange{\pageref{firstpage}--\pageref{lastpage}}
\maketitle

\begin{abstract}
A small fraction of GRBs with available data down to soft X-rays ($\sim0.5$~keV) have been shown to feature a spectral break in the low-energy part  ($\sim$1-10~keV) of their prompt emission spectrum. The overall spectral shape is consistent with optically thin synchrotron emission from a population of 
particles that have cooled on a timescale comparable to the dynamic time to energies that are still much higher than their rest mass energy (marginally fast cooling regime). We consider a hadronic scenario and investigate if the prompt emission of these GRBs can originate from relativistic protons that radiate synchrotron in the marginally fast cooling regime. Using semi-analytical methods, we derive the source parameters, such as magnetic field strength and proton luminosity, and calculate the high-energy neutrino emission expected in this scenario. We also investigate how the emission of secondary pairs produced by photopion interactions and $\gamma\gamma$ pair production affect the broadband photon spectrum. We support our findings with detailed numerical calculations. Strong modification of the photon spectrum below the break energy due to the synchrotron emission of secondary pairs is found, unless the bulk Lorentz factor is very large ($\Gamma \gtrsim 10^3$). Moreover, this scenario predicts unreasonably high Poynting luminosities because of the strong magnetic fields ($10^6-10^7$~G) that are necessary for the incomplete proton cooling. Our results strongly disfavour marginally fast cooling protons as an explanation of the low-energy spectral break in the prompt GRB spectra.
\end{abstract}




\section{Introduction}
Gamma-ray bursts (GRBs) are extremely energetic explosions that release most of their electromagnetic output in $\gamma$-rays within a brief period of time, typically lasting from a fraction of a second to several hundred seconds. GRB light curves are highly variable and consist of several pulses, each of them having a typical width of 10~ms--1~s. The prompt emission is typically observed in the 10~keV--1~MeV energy band \citep{2000P} and its isotropic luminosity can be as high as $10^{54}$ $\rm erg \ s^{-1}$, making GRBs the most luminous objects in the sky \citep[for reviews, see][]{Piran04,KUMAR2015}. 

The radiation mechanism behind the GRB prompt emission is still under debate.
A standard approach adopted to investigate the origin of the prompt emission  is spectral analysis.
This procedure involves fitting empirical functions to the prompt spectra and comparing them with the expectations from different high-energy radiative processes. In many cases, the prompt GRB spectra can be described by a smoothly connected broken power law, known as the ``Band-function'' \citep{Band}.

The fact that the overall spectrum of the GRB prompt emission might be non-thermal had led to the suggestion that the radiation is dominated by synchrotron emission from a power-law distribution of relativistic electrons \citep{1994K,1996SNP}. However, a major criticism of this model
is the predicted low-energy spectral slope. Many prompt emission spectra are represented at low energies by a power law with a hard photon index, namely $dN/d\varepsilon \propto \varepsilon^\alpha$ with $\alpha \sim -1$. However in the standard synchrotron fast cooling model, the spectrum below the peak is expected to have a softer index ($\alpha=-3/2$), making synchrotron radiation a debated process for the interpretation of GRB prompt emission. This problem is known as the synchrotron ``line of death'' \citep{Crider,Preece_2002}. Other problems with a synchrotron interpretation of the prompt emission are the narrow peak-energy distribution and the narrow spectral width of the observed ``Band-function''
as compared with the synchrotron peak, making the suggestion of the synchrotron model highly debatable. 

Variants of the non-thermal electron emission models have been discussed to overcome the ``line-of-death'' problem. One assumption is that the electron synchrotron spectrum is modified, on the low-energy part, by inverse Compton (IC) scattering  in the Klein Nishina regime \citep{derishev2001}. Alternatively, IC scattering of slow cooling electrons on the self-absorbed part of the
synchrotron spectrum was suggested to produce the keV part of the prompt emission spectrum \citep{Panaitescu_2000}. It was also proposed that the electrons should be continuously accelerated over the lifetime of a source in order to reproduce the hardness of most of the observed GRB spectra by the synchrotron self-Compton radiation~\citep{Stern04} . Some other models take into account the electron distribution in pitch angles \citep{Lloyd_Ronning_2002} and the decay of the magnetic field 
over a length scale shorter that the comoving width of the emitting region
\citep{Pe_er_2006,Uhm14} in order to overcome the problem of the electron fast cooling and reproduce the prompt emission spectrum. Synchrotron-self Compton cooling in the Klein Nishina regime for baryon-dominated jets or synchrotron cooling in a  decaying magnetic field for Poynting-dominated jets could lead to low-energy spectra with photon indices consistent with the observed ones, as demonstrated by \cite{2018ApJS..234....3G}. More recently, \cite{2020burges}  performed time-resolved spectral analysis to the prompt spectra of single-pulse GRBs detected with the \emph{Fermi} Gamma-ray Burst Monitor (GBM), while considering a time-dependent synchrotron spectral model with an electron distribution ranging from extremely cooled to extremely uncooled. These authors conclude that the electron synchrotron interpretation of the prompt GRB emission is a feasible option, once time dependence and cooling are properly included.

Motivated by the evidence of inconsistency between  the optically thin synchrotron model and the observations, an alternative,  which contains a thermal component, has been also suggested  \citep{Goodman,1994Thomson}. According to the so-called photospheric model, if the energy dissipation occurs  in the inner part of the outflow, it thermalizes. When the source becomes transparent, at a distance known as the photospheric radius, these photons escape and create a quasi thermal spectrum that peaks at approximately 1 MeV.  If the dissipation occurs near the photospheric region, the thermal photons serve as seeds to relativistic electrons which interact via inverse Compton scattering. \cite{1997LK} proposed that the Comptonization of thermal electrons may be the source of the GRB prompt emission. Moreover \cite{Ghisellini_1999} suggested that the dissipated energy is continuously distributed among all the  thermal  electrons, which may  have a Maxwellian distribution with a normalized temperature  $\Theta$. Consequently, in the presence of continuous energy dissipation and as the flow becomes optically thin, the resulting spectra obtain a non-thermal appearance because of the Comptonization of the quasi-thermal emission by thermal electrons \citep{Meszaros_2000,Pe_er_2005,giannios06}.

Recent studies have argued that observations extending to lower frequencies (optical and soft X-rays) can address the questions regarding the inconsistency of the non-thermal interpretation of the GRB prompt emission, by determining in a more robust way the low-energy photon index and the spectral width of the GRB spectrum. \cite{Oganesyan17}
have performed spectral analysis to GRB spectra during the prompt phase including data from the \emph{Neil Gehrels Swift} X-Ray Telescope (\emph{Swift}/XRT) and \emph{Fermi} GBM \citep{ravasio18,Ravasio_2019}.  \cite{Oganesyan18} extended their analysis  by including data from the \emph{Swift} Burst Alert Telescope (BAT),  and even more recently \citep{Oganesyan19} by including optical observations from the \emph{Swift} UltraViolet and Optical Telescope (UVOT) \citep{Roming_2005} and from ground based robotic telescopes \citep{Lipunov,2005NewA...10..409B,Klotz_2009,2017AstBu..72...81B}. All these studies have shown that the spectrum below 10 keV does not lie on the extrapolation of the low-energy power law of the ``Band function'', but shows a spectral break at around a few keV.  \citet{Oganesyan19} have adopted another phenomenological function to fit the overall prompt GRB spectral energy distribution (SED) that consists of three power laws joined at two energies, $E_{\rm c,obs}$ and $E_{\rm pk,obs}$. Below $E_{\rm c,obs}$ the average photon spectral index is found to be $\alpha_{1}\sim - 2/3$, between $E_{\rm c,obs}$ and $E_{\rm pk,obs}$ is $\alpha_{2}\sim - 3/2$, and above $E_{\rm pk,obs}$ the photon index $\beta$ becomes close to $-2.3$ or slightly steeper.

Relying on these spectral fitting results, \cite{Oganesyan19} reproduced the GRB spectrum  with an electron synchrotron model. They showed that synchrotron radiation is still a viable mechanism for the GRB prompt emission,
if the electron cooling is not complete, but stops at an electron Lorentz factor $\gamma_{\rm c}$\footnote{$\gamma_{\rm c}$ corresponds to the synchrotron  energy $E_{\rm c,obs}$.} that is comparable to the minimum electron Lorentz factor $\gamma_{\rm m}\gg 1$. This idea of  marginally fast cooling electrons with $\gamma_{\rm m}/\gamma_{\rm c}=\mathcal{O}(10)$, which has been thoroughly discussed in earlier works  \citep{kumar08,Daigne11,2013Beniamini-Piran,2014Beniamini-piran}, is a possible solution to the inconsistency between the expected and the measured photon indices found in previous works. 
\cite{Oganesyan19} generated successfully electron synchrotron spectra and from the observables managed to compute some source parameters as a function of the bulk Lorentz factor, such as the magnetic field of the source, the distance from the central region, the electron distribution power law index and the number of relativistic emitting electrons. 

\cite{ghisselini20} also examined the idea that synchrotron radiation  from marginally fast cooling particles is able to reproduce GRB prompt emission. They calculated the proper values of source radii and magnetic fields expected in the case of a leptonic scenario, where the radiating particles are electrons, and showed that the results contradicted the general ideas of GRB phenomenology. More specifically, the short variability timescales,  often seen in the prompt emission of GRBs, indicate that the emitting region must be compact and located at relatively short distances ($R_{\gamma} \sim 10^{14}-10^{16}$~cm) from the central engine. Assuming that the magnetic field of the jet is decaying with distance from the central engine, a compact emitting region should also contain a strong magnetic field. According to the calculations of \cite{ghisselini20}, the synchrotron radiation  from marginally fast cooling electrons is able to reproduce the GRB prompt emission, if the radius of the emitting region is $R_\gamma\gtrsim10^{16}$ cm and the comoving magnetic field strength is $B\lesssim1$ G; both values suggest that the minimum variability timescale should be much larger than the one observed. As a way out of these inconsistent results, \cite{ghisselini20} proposed that the prompt emission originates from marginally fast synchrotron cooling protons.

The aim of this project is to extend the work of \cite{Oganesyan19} and \cite{ghisselini20} by considering a hadronic scenario for the prompt emission. More specifically, our working hypothesis is that synchrotron radiation from a population of relativistic protons  gives rise to the observed prompt GRB emission. Using the best-fit values obtained by \cite{Oganesyan19} for the low-energy spectral break $E_{\rm c, obs}$ of the prompt GRB spectrum, the flux at that energy, $F_{\rm c}$ (in units of mJy), and the ratio of the two break energies, $E_{\rm pk, obs}/E_{\rm c, obs}$,  we estimate analytically the source parameters in the proton synchrotron scenario for about two dozens GRBs as a function of the bulk Lorentz factor.
We then numerically compute the photon spectra while including all relevant radiative processes besides proton synchrotron radiation. In particular, our numerical calculations take into account $\gamma \gamma$ pair production processes as well as interactions of protons with radiation (photohadronic interactions), namely photopair (Bethe-Heitler) pair production and photomeson production processes. All these processes are responsible for the injection of ultra-relativistic pairs  in the emitting region. These can efficiently radiate their energy through synchrotron radiation, thus shaping the overall GRB spectrum. Another result of the photohadronic interactions is the production of photons with energies well above the peak synchrotron energy. Depending on source parameters, these photons can be attenuated via $\gamma \gamma$ pair production, thus producing even more pairs and modifying the GRB prompt spectrum \citep[see also][]{2014MNRAS.442.3026P}. Our goal is to test the role of these additional physical processes in shaping the overall GRB spectrum, and identify parameter regimes where the proton synchrotron scenario for the prompt emission is valid. Lastly, we complement our analysis by calculating the accompanying high-energy neutrino signal expected in this scenario for all plausible parameter sets.


This work is structured as follows. In Section \ref{Sec:2} we derive the model parameters of the problem. Continuously, in Section \ref{Sec:3} we present the numerical code that we utilize in order to construct the GRB photon spectra. In Section \ref{Sec:4} we calculate semi-analytically the neutrino fluxes of all the GRBs of our sample and discuss the significance of the photohadronic processes and of $\gamma \gamma$ pair production, by showing some analytical results  for the parameter values of one specific example. Afterwards we show the numerical results for the same parameter values. We conclude in Section \ref{Sec:6}
with a summary and a discussion of our results. Throughout this study we use $H_{\rm 0} = 69.32$ $\rm km \
Mpc^{-1}\ s^{-1}$, $\Omega_{\rm M} = 0.29$, $\Omega_{\rm \Lambda} = 0.71$ \citep{wmap9}. We use published values for GRB redshifts, while we adopt $z=2$ for GRBs without measured redshift.

\section{Determination of model parameters} 
\label{Sec:2}

As we mentioned above, our first goal is to  reexamine the general synchrotron model for the GRB prompt emission, following the analysis of \cite{Oganesyan19} and \cite{ghisselini20}, in the case of a hadronic scenario and determine the regime in the parameter phase space in which this scenario is viable. For the analytical calculations presented in this section, we make the implicit assumption that  synchrotron radiation dominates the proton energy losses inside the source and is responsible for the prompt GRB emission.

We  assume that at a distance $ R_{\gamma}$ from the central engine  of the GRB, relativistic protons are injected inside a spherical region   that moves with a bulk Lorentz factor $\Gamma$.  This spherical region can be thought of as the shell of the shocked ejecta in the internal shock GRB model. It has got a comoving width $r_{\rm b}=R_{\gamma}/\Gamma$ and contains a tangled magnetic field of comoving strength $B$.

Relativistic protons are injected in the emitting region after being accelerated only once into a power law distribution of spectral index $p$, starting from a minimum Lorentz factor $\gamma_{\rm m}$ up to a maximum value $\gamma_{\rm max}$. The proton injection rate (per unit volume) can be written as
\begin{equation}
    Q_{\rm p}(\gamma,t)=Q_{0} \gamma^{-p} H(\gamma-\gamma_{\rm m}) H(\gamma_{\rm max}-\gamma)H(t),
    \label{eq:Qp}
\end{equation}
where $H(x)$ is the Heaviside function and   
$\gamma_{\max}=\min(\gamma_{\rm H}, \gamma_{\rm eq})$. Here, $\gamma_{\rm eq}$ is the Lorentz factor where the synchrotron loss timescale equals the acceleration timescale $t_{\rm acc}= \eta r_{\rm g}/c = \eta m_{\rm p} \gamma c/qB$ with $\eta \ge 1$, and $\gamma_{\rm H}$ is the Lorentz factor of protons with $r_{\rm g}= r_{\rm b}$ \citep{Hillas}. Here, $q, m_{\rm p}$ are the charge and mass of the proton, and $c$ is the speed of light.

The injection rate of eq. (\ref{eq:Qp}) translates also to an injection luminosity of relativistic protons in the comoving frame as 
\begin{equation}
    L_{\rm p}=\frac{4 \pi r_{\rm b}^{3}}{3} m_{\rm p} c^{2} Q_{\rm 0} \int^{\gamma_{\rm max}}_{\gamma_{\rm m}} \gamma^{-p+1} d\gamma.
    \label{eq:Lp}
\end{equation}
The proton injection luminosity $L_{\rm p}$ can in turn be used to define the proton injection compactness, $\ell_{\rm p}$, as
\begin{equation}
    \ell_{\rm p}=\frac{L_{\rm p}  \sigma_{\rm T}}{4 \pi r_{\rm b}  m_{\rm p} c^{3}} \cdot
    \label{eq:lp}
\end{equation}

Upon entering the source, the relativistic protons interact with the magnetic field and any soft photons present, creating radiation and secondary particles, through proton synchrotron radiation and photohadronic (i.e., photomeson and photopair) interactions respectively. In order to make the free parameters as few as possible, we assume that the injection of primary relativistic electrons has a negligible contribution to the photon emission. Because of this, the main target photons for photomeson and photopair processes are the proton synchrotron photons.

According to \cite{Oganesyan19}, for most of the analysed spectra of that work, spectral fits with a synchrotron function returned a well constrained cooling energy $E_{\rm c,obs}$ and ratio  $\gamma_{\rm m}/\gamma_{\rm c}$. These in turn constrain the peak
energy $E_{\rm pk,obs}$ of the prompt GRB spectrum
\begin{equation}
E_{\rm pk,obs}=\left(\frac{\gamma_{\rm m}}{\gamma_{\rm c}}\right)^2 E_{\rm c,obs}.  
\label{eq:epec}
\end{equation}
These quantities 
in addition to the flux  at the cooling energy $F_{\rm c}$  and  the variability timescale of the GRB prompt emission $t_{\rm \gamma,obs}$, are four observables that can lead us to the source characteristics needed to explain the observed spectra as proton synchrotron radiation. 

The observed proton synchrotron spectrum can be described by the following free parameters: the magnetic field strength $B$, the minimum Lorentz factor of the proton distribution $\gamma_{\rm m}$, the power-law slope $p$ of the proton distribution, the distance from the central engine $R_{\rm \gamma}$ (or comoving size of the emitting region $r_{\rm b}$), the comoving proton luminosity $L_{\rm p}$, and the bulk Lorentz factor $\Gamma$. The latter parameter cannot be determined from the four available observables. Hence, we express all quantities as a function of the bulk Lorentz factor. In the marginally fast cooling scenario considered here, the power-law slope $p$ is related to the photon index $\beta$ above the peak energy of the prompt photon spectrum, as $\beta = -p/2-1$. \cite{Oganesyan19} could not constrain the high-energy part of the spectrum because of the lack of \emph{Fermi}-GBM data, except for one GRB. Motivated by the results of spectral analysis performed on large GRB samples with empirical models  (e.g. \cite{2011Nava}, \cite{Goldstein_2013}, \cite{Gruber_2014}, we use $\beta=-2.3$  throughout this work.
 
We express the three observables ($E_{\rm pk,obs}$, $E_{\rm c, obs}$, and $F_{\rm c}$), as a function of the bulk Lorentz factor and the other source parameters, that are required to give rise to proton synchrotron spectra with the same observational characteristics
 \begin{equation}
     E_{\rm pk,obs}=\frac{m_{\rm e}}{m_{\rm p}} \frac{3 q h B \gamma_{\rm m}^{2}}{4 \pi m_{\rm e} c} \frac{\Gamma}{1+z}
     \label{eq:Ep}
 \end{equation}
 \begin{equation}
     E_{\rm c,obs}=\left(\frac{m_{\rm p}}{m_{\rm e}}\right )^{5} \frac{27 \pi q h m_{\rm e} c (1+z)}{\sigma_{\rm T}^{2} B^{3} t_{\gamma, \rm obs}^{2} \Gamma (1+Y)^{2}}
     \label{eq:Ec}
 \end{equation}

  \begin{equation}
     t_{\rm \gamma,obs}=\frac{R_{\rm \gamma} (1+z)}{c \Gamma^{2}},
     \label{eq:tvar}
 \end{equation}
where $z$ and $d_{\rm L}$ are the redshift and luminosity distance, $Y=U_{\gamma}/U_{\rm B}$ is the Compton parameter, and $U_{\gamma}$ is the photon energy density. In our analysis we assume that $Y \rightarrow 0$, since the losses via IC scattering are negligible for the parameter values in this hadronic scenario (for details, see Appendix~\ref{addxc}).

The last observable, $F_{\rm c}$, is expressed through the bolometric flux $F_{\gamma}$ as 
\begin{equation}
F_{\rm c}= F_{\rm \gamma} \left(\frac{E_{\rm c,obs}}{h}\right)^{-1} \left(\frac{3}{4}+2\sqrt{\frac{E_{\rm pk,obs}}{E_{\rm c,obs}}} -2 +\frac{2}{p-2} \sqrt{\frac{E_{\rm pk,obs} }{E_{\rm c,obs}}}\right)^{-1}\cdot
\label{eq:Fc}
\end{equation}
From eq.~\ref{eq:Ec} we find that the comoving magnetic field of the source is expressed as
\begin{eqnarray}
    B&=&\frac{1}{\Gamma^{1/3}} \left(\frac{27 \pi q h m_{\rm e} c (1+z)}{\sigma_{\rm T}^{2} t_{\rm \gamma,obs}^{2}   E_{\rm c,obs}} \right)^{1/3} \left(\frac{m_{\rm p}}{m_{\rm e}}\right)^{5/3} \\
    &\simeq&6\times10^6~{\rm G} \left(\frac{\Gamma}{300}\right)^{-1/3}\!\!\left(\frac{t_{\gamma, \rm obs}}{1~{\rm s}}\right)^{-2/3}\left(\frac{E_{\rm c,obs}}{10~{\rm keV}}\right)^{-1/3}\left(\frac{1+z}{3}\right)^{1/3}
    \label{eq:b}
\end{eqnarray}
The minimum proton energy $\gamma_{\rm m}$ is computed from eq. \ref{eq:epec}, taking into account eq. \ref{eq:Ep}
\begin{eqnarray}
    \gamma_{\rm m}&=&\sqrt{ \left(\frac{\gamma_{\rm m}}{\gamma_{\rm c}}\right)^{2} \frac{4 \pi m_{\rm e} c (1+z)}{ 3 q h} \frac{m_{\rm p}}{m_{\rm e}}  \frac{E_{\rm c,obs}}{B \Gamma}} \\ 
    &\simeq&\!\!\!\!1.3\times10^4 \left(\frac{\zeta}{10}\right)\left(\frac{\Gamma}{300}\right)^{-1/3} \left(\frac{t_{\gamma, \rm obs}}{1~{\rm s}}\right)^{1/3}\left(\frac{E_{\rm c,obs}}{10~{\rm keV}}\right)^{2/3}\left(\frac{1+z}{3}\right)^{1/3},
    \label{eq:gm}
\end{eqnarray}
where $\zeta \equiv \gamma_{\rm m}/\gamma_{\rm c}$.
The distance from the central engine is computed from eq.~\ref{eq:tvar} and reads
\begin{eqnarray}
    R_{\gamma} &=& c \Gamma^{2} t_{\rm \gamma,obs} (1+z)^{-1} \\ 
    &=& 9\times10^{14}~{\rm cm} \left(\frac{\Gamma}{300}\right)^{2} \left(\frac{t_{\gamma, \rm obs}}{1~{\rm s}}\right) \left(\frac{1+z}{3}\right)^{-1},
    \label{eq:rg}
\end{eqnarray}
Accordingly, the size of the emitting region $r_{\rm b}$ is 
\begin{eqnarray}
    r_{\rm b} &=& c \Gamma t_{\rm \gamma,obs}(1+z)^{-1} \\ 
    & =& 3\times10^{12}~{\rm cm} \left(\frac{\Gamma}{300}\right) \left(\frac{t_{\gamma, \rm obs}}{1~{\rm s}}\right) \left(\frac{1+z}{3}\right)^{-1} \cdot
    \label{eq:rb}
\end{eqnarray}
In order to compute the comoving proton luminosity, we assume that the observed luminosity is approximately equal to the injection luminosity of particles. This hypothesis is valid as long as particles are in the marginally fast cooling regime. We therefore may write

\begin{eqnarray}
    L_{\rm p} &\approx& \frac{ L_{\gamma, \rm obs}}{\Gamma^{4}} \simeq 4.6\times10^{42}~{\rm erg\, s^{-1}}\left(\frac{F_{\rm c}}{1~{\rm mJy}}\right)\left(\frac{E_{\rm c,obs}}{10~{\rm keV}}\right)\\ \nonumber 
    &&\left(\frac{\Gamma}{300}\right)^{-4}  \left[\frac{3}{4}+2\left(\frac{\zeta}{10}\right)-2+3.3\left(\frac{\zeta}{10}\right)\right]\cdot
\end{eqnarray}
where $L_{\gamma, \rm obs}=4 \pi d_{\rm L}^{2} F_{\rm \gamma}$ is the bolometric $\gamma$-ray luminosity of the prompt emission in the  observer's frame, and the bolometric flux, $F_{\rm \gamma}$, is computed from eq.~\ref{eq:Fc}.    

All the above source parameters are going to be displayed as a function of the bulk Lorentz factor $\Gamma$ in Sec.~\ref{Sec:4}. The best-fit values (and the $1\sigma$ uncertainties) of the four observables, required for the calculation of source parameters, are taken from a sample of 21 GRBs, shown in \cite{Oganesyan19} (see Table B1 of Appendix B). For GRBs where  analysis is performed in multiple time intervals, we show results for one time interval that corresponds to the brightest flux state, unless stated otherwise. In our analysis we do not include GRB~060814 and GRB~090715B because more than one observables are not constrained (they have upper or lower limits). Other bursts that have upper limits on only one observable have been included in the analysis.

\section{Numerical Code}
\label{Sec:3}
In the previous section we  provided analytical expressions for computing the source parameters in the context of a proton synchrotron model for the GRB prompt emission.
To verify this assumption, we utilize a time-dependent numerical code  \citep[\texttt{ATHE$\nu$A},][]{DM12} that follows the evolution of spatially averaged particle populations inside a homogeneous spherical emitting region. The numerical approach of the problem gives us also the opportunity to extend the work of \cite{ghisselini20} by investigating the contribution of photohadronic interactions and $\gamma\gamma$ pair production to the overall photon spectrum and by computing the associated neutrino flux.

The numerical code computes the  electromagnetic, neutrino, and cosmic-ray (i.e., proton and neutron) fluxes emerging from a single radiation zone in the jet under certain assumptions. These are summarized below. After being injected inside a spherical source with a constant injection rate (as in eq.~\ref{eq:Qp} but with a high-energy exponential cutoff instead of a sharp cutoff at $\gamma_{\max}$), relativistic protons interact with
the magnetic field and any soft photons present,  creating radiation and other secondary particles. Among those, pions and muons decay almost instantaneously into lighter particles (i.e., pairs, electron and muon neutrinos). However,  a fraction of their populations can cool by emitting synchrotron photons before their decay (for more details, see Appendix \ref{apb}). Pion and muon synchrotron cooling can affect both the photon and neutrino spectra \citep{2011Baerwald,PGD14,2015Tamborra}. Neutrons do not typically interact with soft photons before they escape the source (i.e., the source is optically thin to neutron-photon interactions), and neutrinos escape the source without any interactions on a light-crossing time. 

At any given time, there are five stable particle species in the emitting region, namely protons, photons, electron-positron pairs, neutrons\footnote{Neutrons typically escape the source without undergoing neutron-photon interactions, and deposit their energy in the surrounding region once they are transformed back to protons.} and neutrinos. The production and loss rates of these five stable particle species are tracked self-consistently with five time-dependent coupled kinetic equations, which can be written in a compact form as
    \begin{equation}
    \frac{\partial n_{\rm i}}{\partial t}+\frac{n_{\rm i}}{t_{\rm i,esc}}+\mathcal{L}_{\rm i}=Q_{\rm i}.
    \label{kineticeq}
\end{equation}
Here, $n_{\rm i}$ is the differential number density of particle species $i$,  $t_{\rm i,esc}=r_{\rm b}/c$  is the respective escape timescale, which is assumed to be energy-independent for all particles, and $Q_{\rm i}$ and $\mathcal{L}_{\rm i}$ are the injection (source)
and loss (sink) terms, respectively. These terms include the following processes: 
\begin{itemize}
   \item synchrotron radiation for both electrons and protons 
    \item proton-photon pair production (photopair)
    \item proton-photon pion production (photopion)
    \item neutron-photon pion production
    \item pion, kaon, muon and electron synchrotron radiation
    \item synchrotron self-absorption\footnote{Heating of particles due to this process is not included in the calculations, but it is expected to be negligible for the parameters used here.}
    \item electron inverse Compton scattering
    \item photon-photon ($\gamma \gamma$) pair production
    \item electron-positron pair annihilation
\end{itemize}
Summarizing, the numerical code we utilize is the one presented in \cite{Mast2005,DM12} but augmented in a way to include pion, muon and kaon synchrotron cooling. Details about the implementation of the latter can be found in \cite{PGD14}.

\section{Results}
\label{Sec:4}

In this section we adopt the idea that the marginally fast synchrotron cooling can relax the inconsistency between the synchrotron radiation and the harder spectra of some GRBs. We adopt the idea of a hadronic scenario \citep{ghisselini20}, according to which, the proton synchrotron is the dominant radiative process in the emitting source responsible for the prompt GRB emission. Based on this framework, we analytically obtain constraints on the physical parameters of the source (Sec.~\ref{sec:values}). As a next step, we numerically calculate the broadband photon spectra and investigate whether their shape is modified when additional physical processes are taken into account (Sec.~\ref{sec:numerical}).

\subsection{The parameter space of the proton synchrotron model for GRB prompt emission}\label{sec:values}

In order to find some constraints on the physical parameters of the source, such as the magnetic field, the size of the emitting region, the proton luminosity, and the minimum proton Lorentz factor, we adopt the best-fit values (and uncertainties) for several observables from \cite{Oganesyan19} and use the analytical expressions presented in Sec.~\ref{Sec:2}. To compute the uncertainties on the inferred model parameters, we use the python package \texttt{soad}\footnote{\url{https://github.com/kiyami/soad/}}. Using the  methods of \cite{python-soad}, \texttt{soad} searches for a likelihood function that has a maximum at the best-fit value of a measured quantity with asymmetric uncertainties.

\begin{figure*}
    \centering
    	\includegraphics[width=0.45\linewidth]{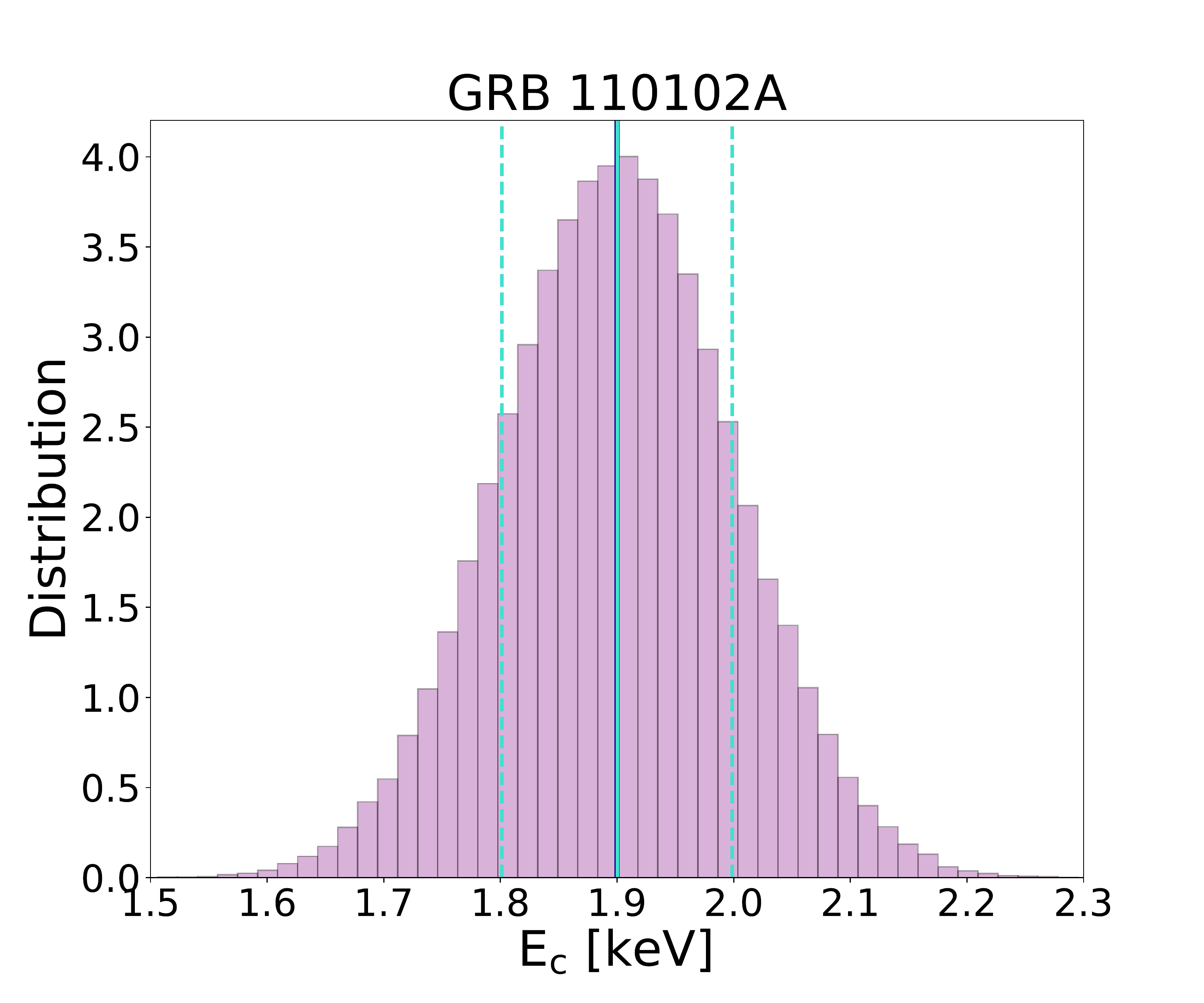}
	\includegraphics[width=0.45\linewidth]{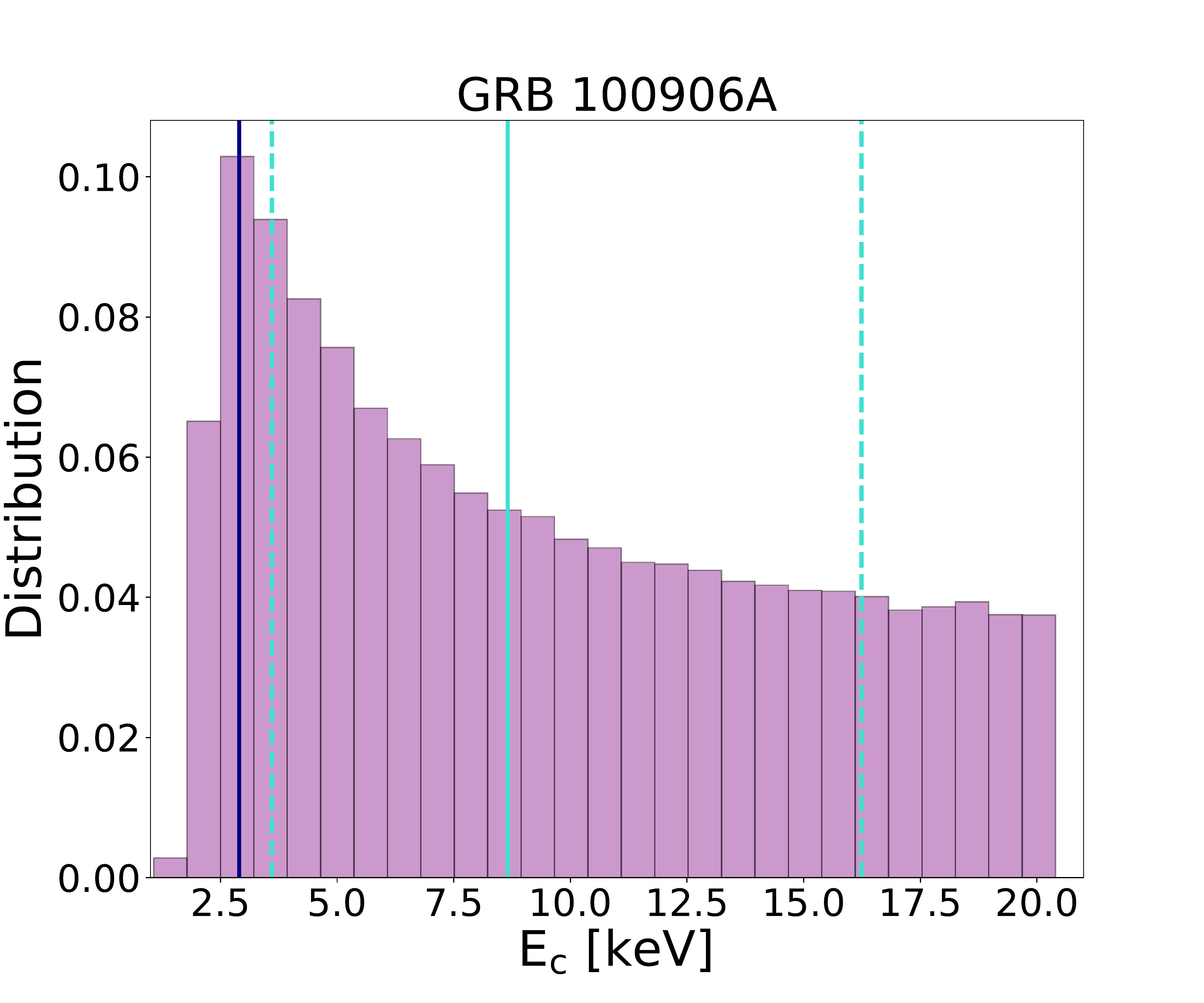}
    \caption{Distribution of $E_{\rm c}$ values for two GRBs created using the python package \texttt{soad} and the best-fit values with their uncertainties from \citet{Oganesyan19}. The solid dark blue and turquoise lines  correspond to  the most probable value and to the median value respectively, while and the turquoise dashed lines indicate the 68\% uncertainty range. }
    \label{fig:uncertaintycode}
\end{figure*}

\begin{figure*}
	\includegraphics[width=0.45\linewidth]{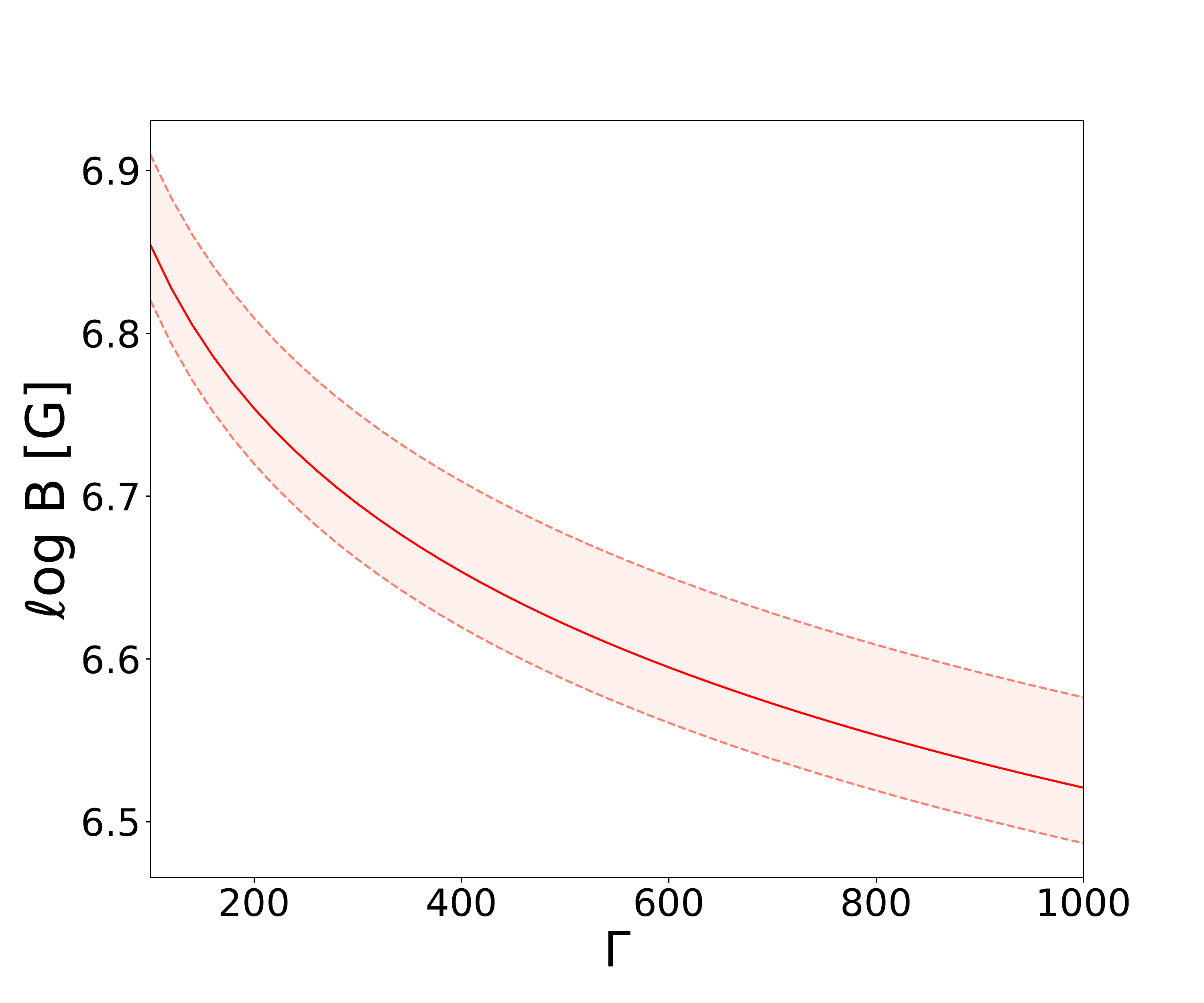}
	\includegraphics[width=0.45\linewidth]{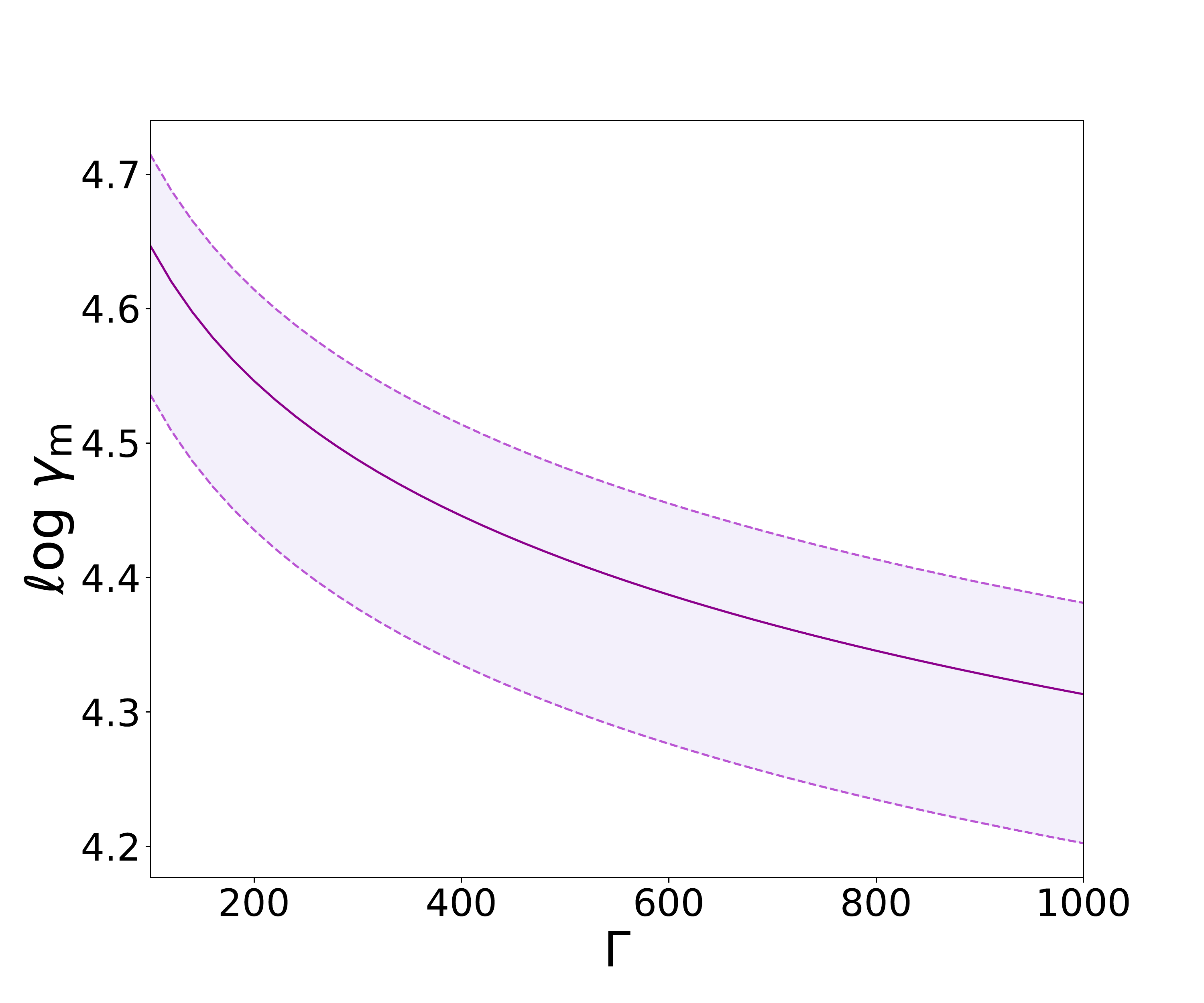}
	\includegraphics[width=0.45\linewidth]{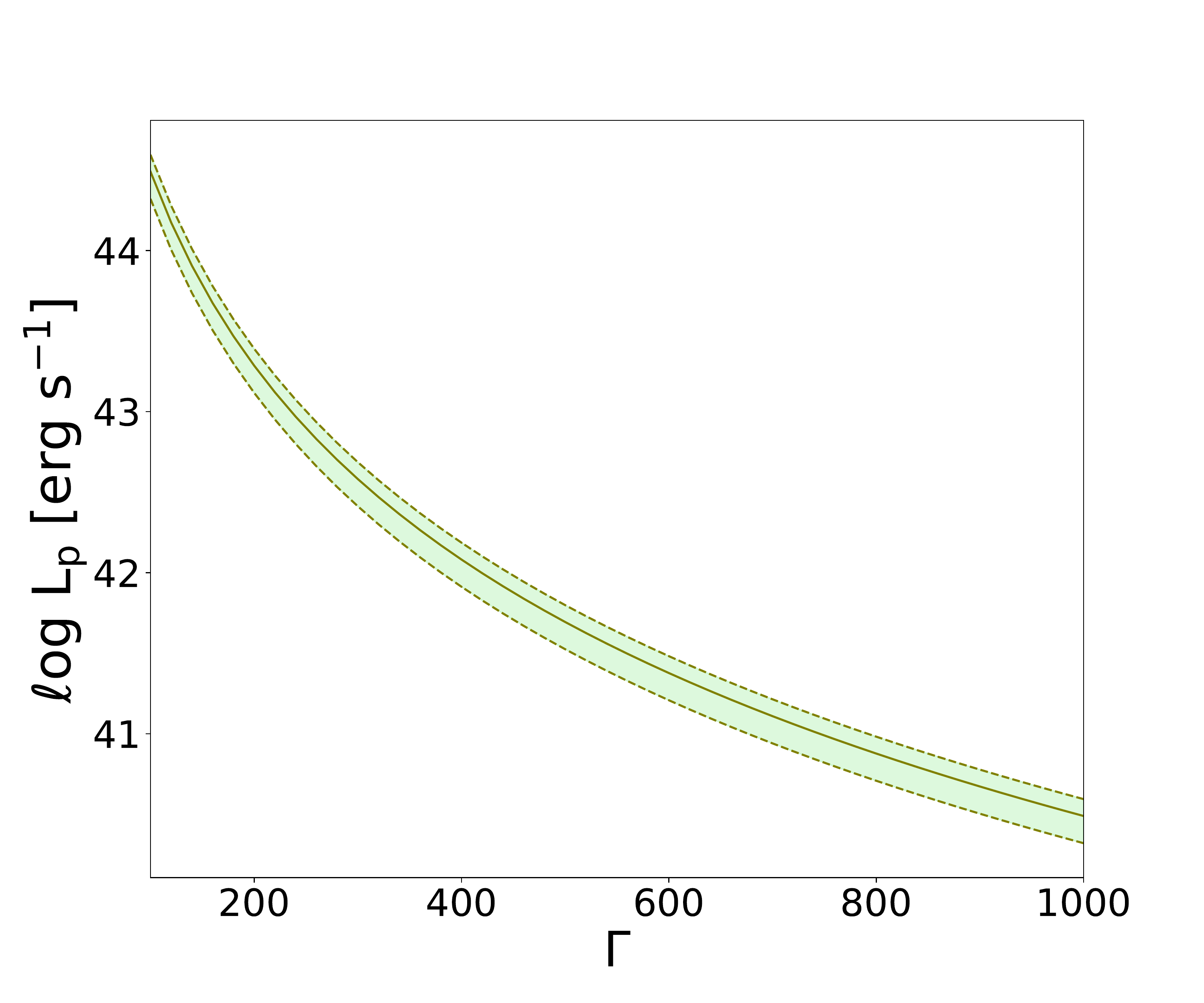}	
	\includegraphics[width=0.45\linewidth]{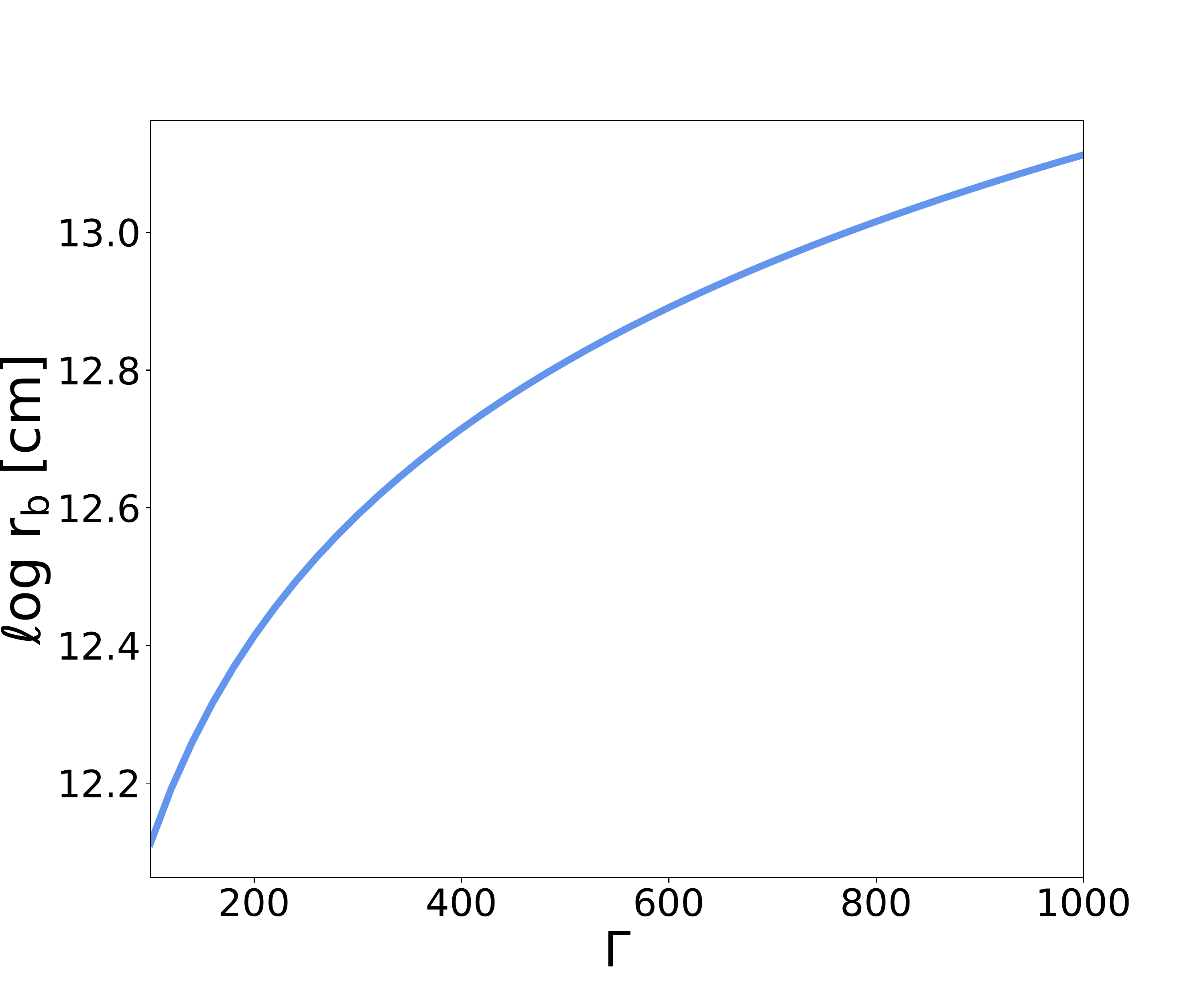}	
    \caption{Inferred parameters of the proton synchrotron model for  GRB~061121 using observables from the time interval with the highest photon flux. From top left and in clockwise order we show the  magnetic field $B$, the minimum proton Lorentz factor $\gamma_{\rm m}$, the source radius $r_{\rm b}$, and the proton injection luminosity $L_{\rm p}$  as a function of the bulk Lorentz factor $\Gamma$. Shaded regions indicate the $1  \sigma$ uncertainties, whenever relevant. All quantities are measured in the comoving frame of the outflow.  
    }
    \label{fig1}
\end{figure*}

As an example, we show in Fig.~\ref{fig:uncertaintycode}, the distributions of $E_{\rm c}$ values for two GRBs of the sample. The $E_{\rm c}$ values of GRB~110102A follow approximately a normal distribution, while the distribution of GRB~100906A is asymmetric with positive skewness.
The turquoise solid and dashed lines shows the median value and the 68\% range of these distributions respectively, while the solid blue line indicates the best-fit value, which can also be interpreted as the most probable value of the distribution. The median value does not coincide with the most probable value, only for a few GRBs in the sample (GRB~081008, GRB~100906A, GRB~121123A, GRB~140206, GRB~151021A) that have very asymmetric distributions for one (or more) observables. In what follows, we report and use the median and 68\% uncertainty range for the model parameters.

All model parameters are plotted as a function of the bulk Lorentz factor of the outflow for a wide range of plausible values, namely $100\le\Gamma\le1000$.  As an illustrative example,  we show in Fig.~\ref{fig1} the values of the comoving magnetic field, the minimum proton Lorentz factor and the proton compactness as a function of $\Gamma$, obtained for GRB~061121 (at redshift $z=1.314$~\citep{061121redshift}). As we have already shown in eqs.~\ref{eq:Fc}, \ref{eq:b}, and \ref{eq:rb}, the values of  $B$,  $\gamma_{\rm m}$ and $L_{\rm p}$ decrease as the bulk Lorentz factor increases. We also show the size of the spherical source $r_{\rm b}$ as a function of $\Gamma$, which is common for all the bursts, since it is computed following eq. \ref{eq:rg} and \ref{eq:rb}, under the assumption that $t_{\gamma}=1$ s for all the GRBs included in the sample. Shorter variability timescales translate to smaller source sizes and higher photon number densities.

\begin{figure}
	\includegraphics[width=\linewidth]{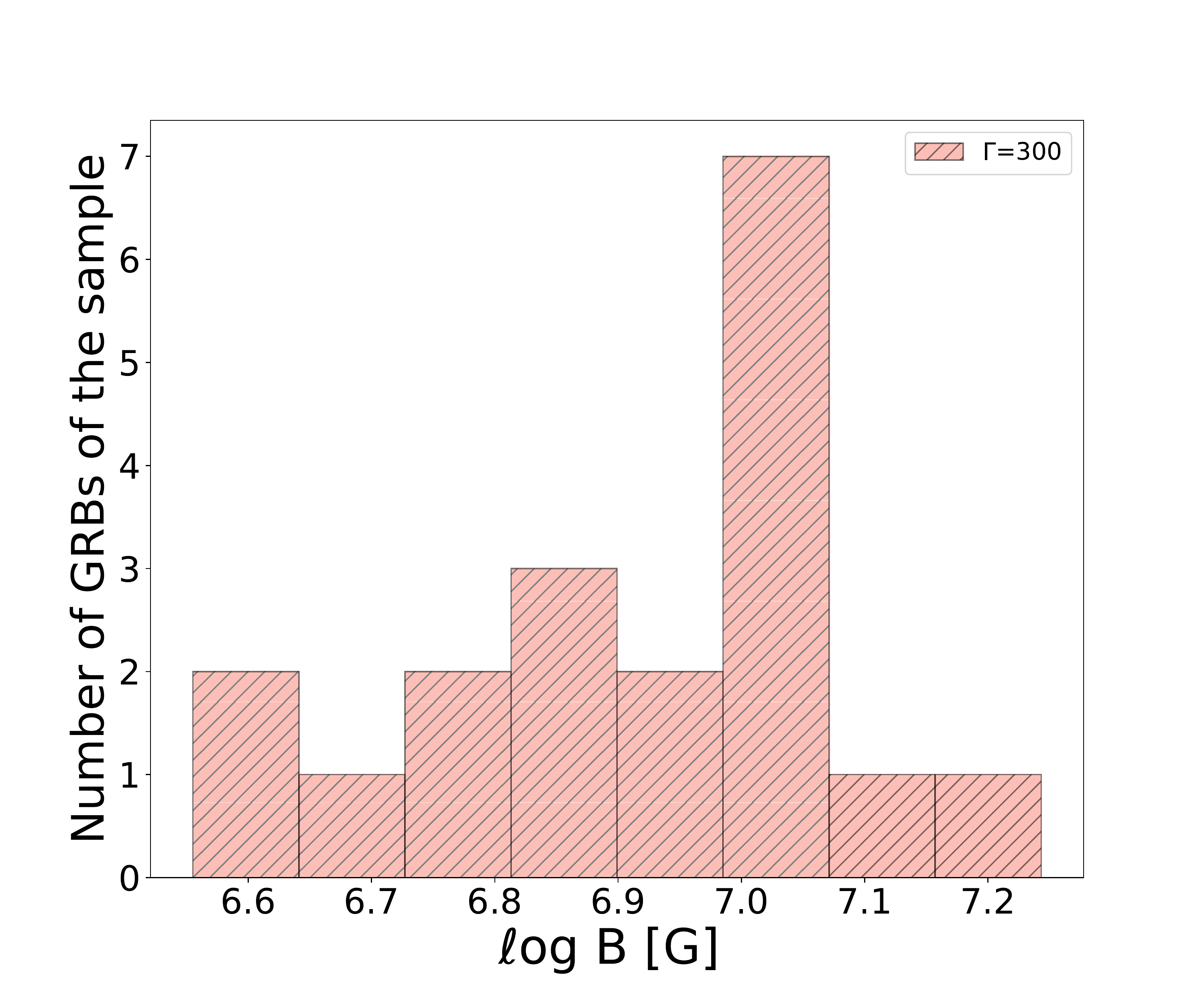}
	\includegraphics[width=\linewidth]{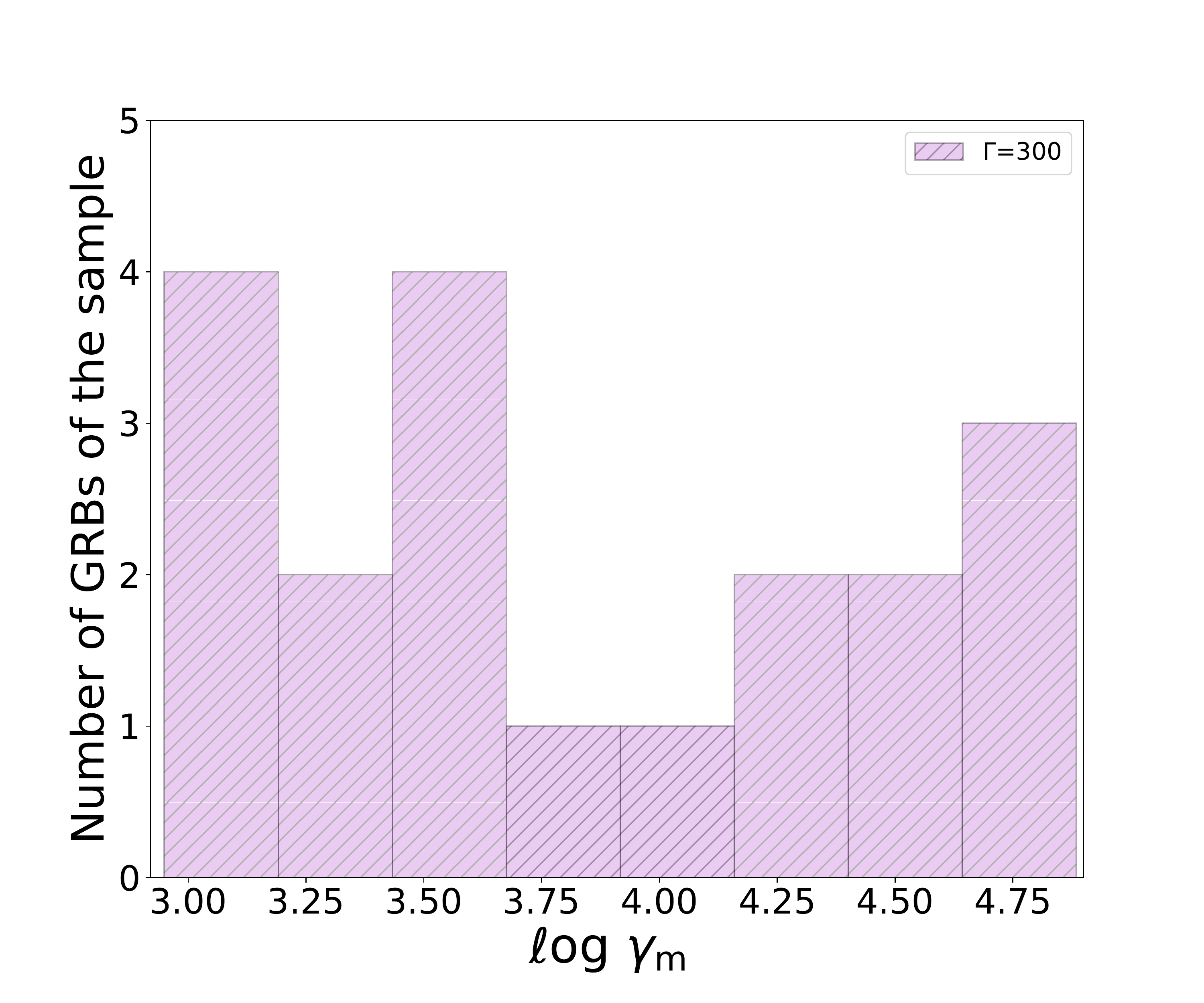}
	\includegraphics[width=\linewidth]{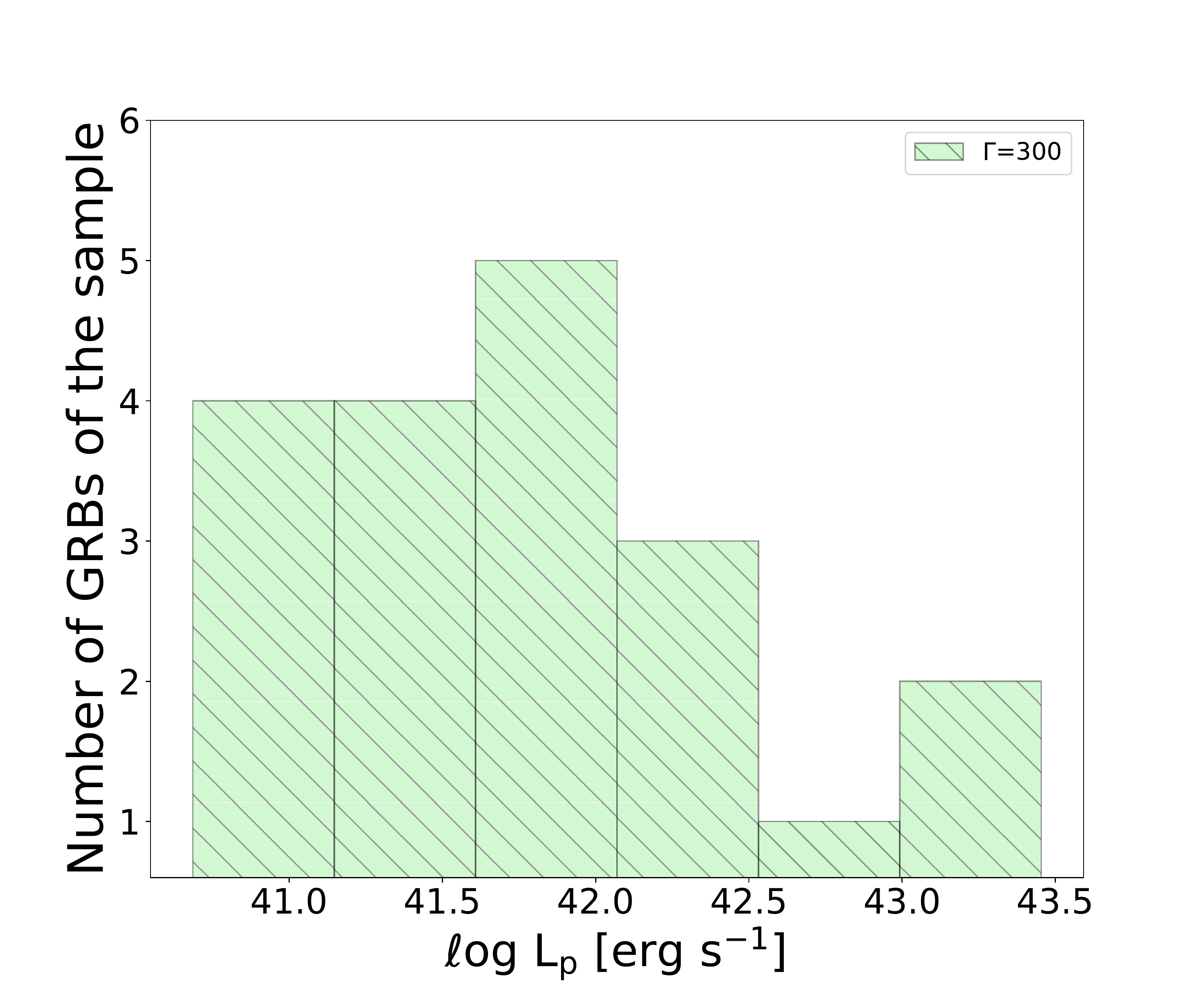}
	\caption{Logarithmic histograms of the inferred $B, \gamma_{\rm m}$, and $L_{\rm p}$ values for our GRB sample and $\Gamma=300$. Results for another choice of $\Gamma$ can be obtained using the scaling relations  $B \propto \Gamma^{-1/3}$ , $\gamma_{\rm m} \propto \Gamma^{-1/3}$ , $L_{\rm p} \propto \Gamma^{-4}$. The values of individual GRBs are listed in Table~\ref{tab:app}.}
    \label{fig3}
\end{figure}
We repeat the same analysis for all GRBs in our sample and report the values of $B, \gamma_{\rm m}$, and $L_{\rm p}$ in Fig.~\ref{fig3} for an indicative value of the bulk Lorentz factor  $\Gamma=300$. Results for another choice of $\Gamma$ can be found using the scaling relations  $B \propto \Gamma^{-1/3}$, $\gamma_{\rm m} \propto \Gamma^{-1/3}$, $L_{\rm p} \propto \Gamma^{-4}$. The parameters for individual GRBs for $\Gamma=300$ and 1000 are listed in Table~\ref{tab:app}.
We find that comoving magnetic fields strengths of the order of $(1-20)\times10^6$~G are required for explaining the spectral break at $\sim$keV energies as result of proton synchrotron cooling. Because of the strong magnetic fields, acceleration of protons to ultra-high energies is not necessary for explaining the keV to MeV part of the GRB prompt spectrum. More specifically, we find that the required minimum  Lorentz factor of the proton distribution, $\gamma_{\rm m}$, ranges between $10^{2.7}$ and $10^{4.8}$. The  proton injection luminosity in the jet comoving frame has a strong dependence on the bulk Lorentz factor (see eq.~\ref{eq:Lp}). As a result, we find that the proton luminosity in the comoving frame spans a wide range of values, namely $10^{38}~{\rm erg \, s}^{-1} \lesssim L_{\rm p} \lesssim 10^{43}$~erg s$^{-1}$. However, the isotropic proton luminosity in the observer's frame is almost independent of $\Gamma$ and the distribution of $L_{\rm p, obs}$ values reflects the distribution of $L_{\gamma, \rm iso}$ in the GRB sample (see  Table~\ref{tab:app}). Finally, we find that the size of the comoving region is $10^{12}~\rm cm< r_{\rm b}\lesssim 2\times10^{13}~\rm cm$. These results are consistent with those of \cite{ghisselini20}, who assumed a power-law particle distribution of minimum lower Lorentz factor $\gamma_{\rm m}=10^{3.5}$, a typical magnetic field $B \sim 10^{6}$~G , and comoving radius of the emitting source $r_{\rm b} \sim 10^{13}$ cm for a fixed variability timescale of $t_{\rm var}=1$ s.

\begin{table*}
\centering
\caption{Median values and 68\% uncertainty ranges (in logarithmic scale) of parameters in the proton synchrotron model for the GRBs analyzed by \citet{Oganesyan19} for $\Gamma=300$. Values enclosed in parentheses are obtained for $\Gamma=1000$.}
\begin{threeparttable}
\begin{tabular}{l c c c c }
\hline \hline
{Source} & {$\rm B$} (G)  & {$\gamma_{\rm m}$}  & {$ \rm L_{\rm p}$ }  (erg s$^{-1}$) &  {$ \rm L_{\rm p,obs}$\tnote{a} }  (erg s$^{-1}$)  \\  
\hline  \smallskip  
{GRB 060510B} & $7.07_{-0.03}^{+0.02}$~($6.89_{-0.02}^{+0.02}$) & $3.51_{-0.19}^{+0.14}$~($3.34_{-0.19}^{+0.14}$) & $42.21_{-0.21}^{+0.16}$~($40.12_{-0.21}^{+0.16}$) & $52.12_{-0.21}^{+0.16}$ \\   \smallskip
{GRB 061121} & $6.61_{-0.02}^{+0.02}$~($6.43_{-0.01}^{+0.02}$) & $>4.46$~($>4.28$ ) &$>43.45$~($>41.35$) & $>53.35$ \\   \smallskip
{GRB 070616} & $7.16_{-0.04}^{+0.06}$~( $6.95_{-0.04}^{+0.05}$) & $3.14_{-0.21}^{+0.25}$~($2.97_{-0.21}^{+0.24}$) & $41.17_{-0.25}^{+0.27}$~($38.08_{-0.25}^{+0.27}$) & $51.08_{-0.25}^{+0.27}$\\   \smallskip
{GRB 080928} & $7.04_{-0.04}^{+0.04}$~( $6.87_{-0.03}^{+0.03}$) & $3.23_{-0.25}^{+0.27}$~($3.06_{-0.24}^{+0.27}$) & $41.39_{-0.27}^{+0.29}$~($39.29_{-0.27}^{+0.29}$) & $51.29_{-0.27}^{+0.29}$\\   \smallskip
{GRB 081008} & $6.86_{-0.09}^{+0.09}$~($6.69_{-0.08}^{+0.08}$) & $4.25_{-0.41}^{+0.35}$~($4.08_{-0.41}^{+0.34}$) & $42.59_{-0.45}^{+0.40}$~($40.49_{-0.44}^{+0.39}$) & $52.49_{-0.44}^{+0.40}$\\   \smallskip
{GRB 100906A} & $6.78_{-0.09}^{+0.13}$~($6.61_{-0.09}^{+0.12}$) & $2.94_{-0.73}^{+0.34}$~($2.7_{-0.72}^{+0.33}$) & $41.90_{-0.88}^{+0.47}$~($39.80_{-0.88}^{+0.46}$) & $51.80_{-0.86}^{+0.47}$\\   \smallskip
{GRB 110102A} & $7.01_{-0.08}^{+0.08}$~($6.84_{-0.01}^{+0.01}$) & $>4.39$~($>4.22$) & $>42.66$~($>40.57$) & $>52.57$\\   \smallskip
{GRB 110119A} & $7.03_{-0.02}^{+0.02}$~($6.85_{-0.015}^{+0.016}$) & $3.46_{-0.08}^{+0.09}$~($3.29_{-0.07}^{+0.09}$) & $41.23_{-0.09}^{+0.09}$~($39.14_{-0.09}^{+0.09}$) & $51.14_{-0.09}^{+0.09}$\\   \smallskip
{GRB 110205A} & $6.89_{-0.03}^{+0.04}$~($6.71_{-0.03}^{+0.04}$) & $3.73_{-0.21}^{+0.25}$~($3.56_{-0.21}^{+0.25}$) & $42.29_{-0.24}^{+0.26}$~($40.20_{-0.24}^{+0.26}$) & $52.20_{-0.24}^{+0.26}$\\   \smallskip
{GRB 111103B} & $7.24_{-0.03}^{+0.041}$~($7.06_{-0.03}^{+0.04}$) & $3.17_{-0.12}^{+0.12}$~($3.00_{-0.12}^{+0.12}$) & $41.00_{-0.15}^{+0.14}$~($38.90_{-0.15}^{+0.14}$) & $50.90_{-0.15}^{+0.14}$\\   \smallskip
{GRB 111123A} & $7.03_{-0.01}^{+0.01}$~($6.86_{-0.01}^{+0.01}$) & $3.57_{-0.07}^{+0.08}$~($3.39_{-0.07}^{+0.08}$) & $41.94_{-0.08}^{+0.09}$~($39.85_{-0.08}^{+0.09}$) & $51.85_{-0.08}^{+0.09}$\\   \smallskip
{GRB 121123A} & $6.39_{-0.08}^{+0.15}$~($6.22_{-0.08}^{+0.15}$) & $3.98_{-0.33}^{+0.30}$~($3.80_{-0.33}^{+0.30}$) & $42.34_{-0.48}^{+0.40}$~($40.25_{-0.48}^{+0.40}$) & $52.25_{-0.48}^{+0.39}$\\   \smallskip
{GRB 121217A} & $6.92_{-0.02}^{+0.03}$~($6.75_{-0.02}^{+0.03}$) & $3.98_{-0.24}^{+0.27}$~($3.80_{-0.24}^{+0.27}$) & $42.08_{-0.24}^{+0.28}$~($40.00_{-0.24}^{+0.28}$) & $52.00_{-0.24}^{+0.28}$\\   \smallskip
{GRB 130514A} & $7.05_{-0.02}^{+0.03}$~($6.87_{-0.02}^{+0.03}$) & $3.28_{-0.09}^{+0.10}$~($3.11_{-0.09}^{+0.10}$) & $41.57_{-0.12}^{+0.12}$~($39.48_{-0.12}^{+0.12}$) & $51.48_{-0.12}^{+0.12}$\\   \smallskip
{GRB 130907A} & $6.82_{-0.04}^{+0.04}$~($6.64_{-0.04}^{+0.04}$) & $2.50_{-0.15}^{+0.13}$~($2.33_{-0.15}^{+0.13}$) & $40.66_{-0.43}^{+0.25}$~($38.57_{-0.43}^{+0.25}$) & $50.57_{-0.43}^{+0.25}$\\   \smallskip
{GRB 140108A} & $6.77_{-0.01}^{+0.01}$~($6.59_{-0.01}^{+0.01}$) & $4.74_{-0.02}^{+0.02}$~($4.56_{-0.02}^{+0.02}$) & $42.50_{-0.04}^{+0.04}$~($40.41_{-0.04}^{+0.04}$) & $52.41_{-0.03}^{+0.04}$\\   \smallskip
{GRB 140206A} & $6.86_{-0.04}^{+0.11}$~($6.68_{-0.04}^{+0.11}$) & $>4.43$~($>4.26$) & $>43.08$~($>41.00$) & $>53.00$\\   \smallskip
{GRB 140512A} & $6.70_{-0.02}^{+0.03}$~($6.52_{-0.02}^{+0.03}$) & $4.44_{-0.25}^{+0.29}$~($4.26_{-0.25}^{+0.29}$) & $41.87_{-0.26}^{+0.29}$~($39.78_{-0.26}^{+0.29}$) & $51.78_{-0.26}^{+0.29}$\\   \smallskip
{GRB 151021A} & $7.02_{-0.11}^{+0.10}$~($6.85_{-0.11}^{+0.10}$) & $3.02_{-0.35}^{+0.31}$~($2.84_{-0.35}^{+0.31}$) & $41.63_{-0.45}^{+0.40}$~($39.54_{-0.45}^{+0.40}$) & $51.54_{-0.45}^{+0.40}$\\ 
\hline 
\end{tabular}
\begin{tablenotes}
\item[a] The proton luminosity in the observer's frame is independent of $\Gamma$. 
\item For GRBs with multiple time intervals analyzed, we report values for the interval of peak flux. In three cases, the minimum Lorentz factor and the proton luminosity are quoted as lower limits, because one of the observables ($\gamma_{\rm m}/\gamma_{\rm c}$) that is required in order to compute them, is given as a lower limit. 
\end{tablenotes}
\end{threeparttable}
\label{tab:app}
\end{table*}

\subsection{The role of photohadronic processes and $\gamma \gamma$ pair production in GRB prompt photon spectra}\label{sec:numerical}

Relativistic protons can become targets to their own radiation, after interacting with the strong magnetic field of the source and emitting synchrotron photons. Before we proceed with fully numerical calculations of the emerging electromagnetic radiation of the GRB prompt phase, we estimate semi-analytically the photohadronic efficiency and $\gamma \gamma$ opacity of the emitting region. 

\begin{figure}
    \centering
    \includegraphics[width=\linewidth]{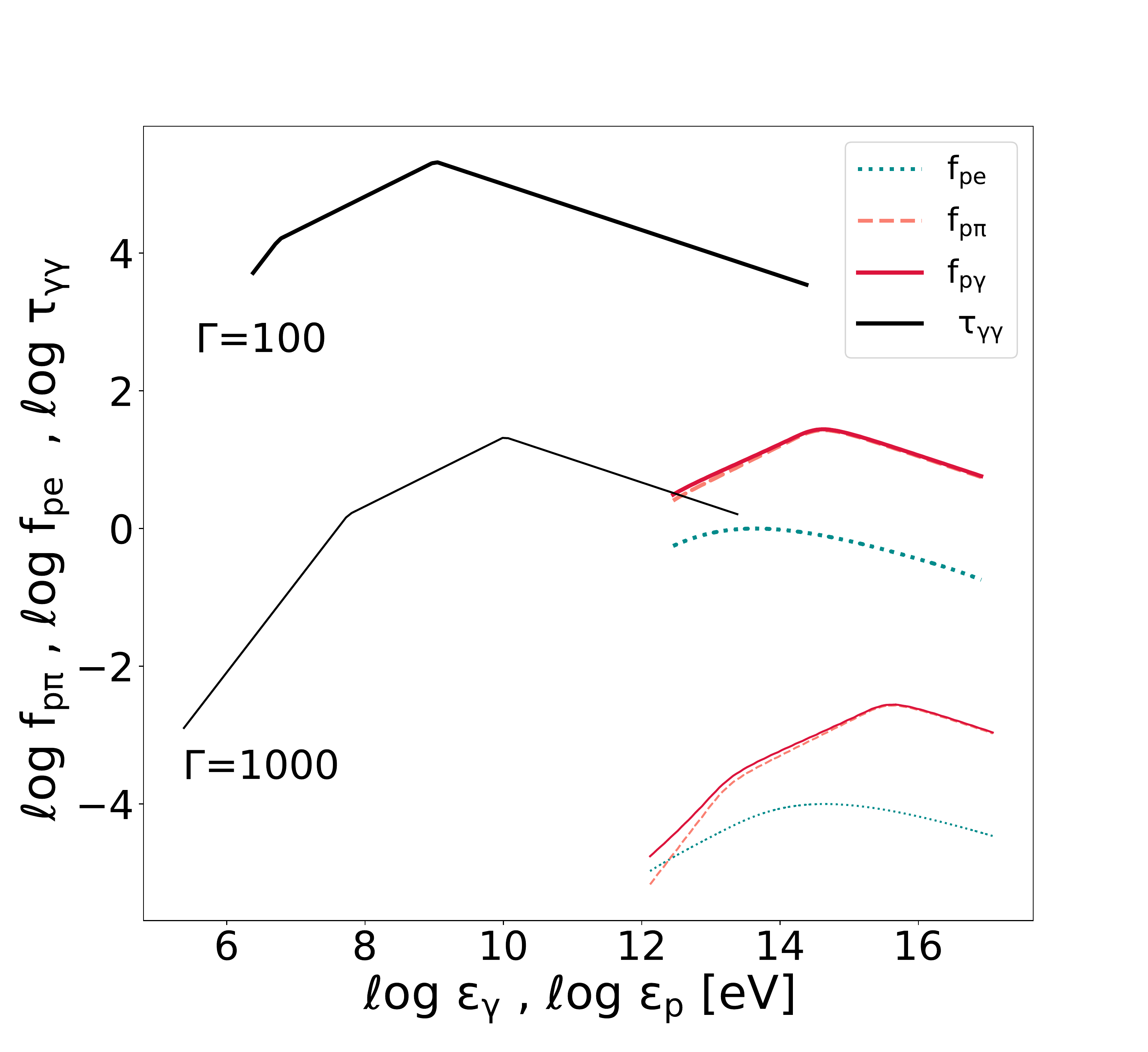}
    \caption{The dimensionless fractional energy loss rate of relativistic protons due to photopion production  (dashed orange lines), photopair production (dotted cyan lines) and the sum of the two processes (solid red lines) as a function of the comoving proton energy. The opacity of the emitting region to $\gamma \gamma$ pair production is also plotted as a function of the comoving photon energy (solid black lines). Thick and thin lines show results for $\Gamma=100$ and 1000, respectively. Displayed results are for GRB~$061121$.
    }
    \label{fig:fboth}
\end{figure}

Fig.~\ref{fig:fboth} shows the photopion production efficiency $f_{\rm p \pi}=t^{-1}_{\rm p \pi}r_{\rm b}/c$ (see eq.~\ref{eq.fpg}
and the photopair production efficiency $f_{\rm p e}=t^{-1}_{\rm p e}r_{\rm b}/c$ (see eq.~\ref{eq.fpe}) as a function of the comoving proton energy $\varepsilon_{\rm p}$ for $\Gamma=100$ (thick lines) and 1000 (thin lines). Solid red lines show the total efficiency of photohadronic interactions, defined as $f_{\rm p\gamma}\equiv f_{\rm p\pi}+f_{\rm pe}$. Proton losses are dominated by photopion production, except for $\Gamma=1000$ where photopair production becomes more important at the lowest energies. The decrease of $f_{\rm p\gamma}$ with energy is a direct outcome of the $\Delta$-resonance approximation for the photopion cross section used here. Consideration of multi-pion production occurring for interaction energies above that of the $\Delta$ resonance would lead to an increase of photopion efficiency with proton energy or saturation \citep[see e.g.,][]{Murase:2008mr, 2014JHEAp...3...29D}. The dependence of $f_{\rm p\gamma}$ on the bulk Lorentz factor for a given proton energy is strong (i.e., $\propto \Gamma^{-4}$) and reflects the dependence of the photon number density on $\Gamma$ (see also eq.~\ref{eq.nx}).

High-energy photons that are produced by the power-law distribution of synchrotron-cooled protons and/or by photohadronic interactions may be absorbed by the isotropic synchrotron photon field inside the emitting region. Therefore, it is important to estimate the photon opacity of the source. We compute the optical depth $\tau_{\gamma \gamma}$ by using eq.~\ref{eq.tau} for the parameter values of GRB~061121 and overplot the solution in Fig.~\ref{fig:fboth} for two values of $\Gamma$ (solid black lines). For most values of $\Gamma$, we find that  the source is optically thick to $\gamma \gamma$ pair production ($\tau_{\gamma \gamma}>>1$) for a wide range of comoving photon energies. Only for $\Gamma=1000$ (thin black line) we find $\tau_{\gamma \gamma} \lesssim 1$ for almost all photon energies. Therefore, we conclude that, for parameters that correspond to lower bulk Lorentz factors, $\gamma \gamma$ photon absorption is unavoidable and the produced electron-positron pairs might affect the overall spectrum of the GRB prompt emission.

\subsection{Prompt photon \& neutrino spectra}
\subsubsection{Semi-analytical results}\label{sec:semianalytic}
GRBs are candidates of proton acceleration to ultra-high energies  and, therefore, potential sources of high-energy neutrino emission \citep{1995Vietri,Waxman_1997,Waxman98}. After investigating the role of the photohadronic interactions in the prompt emission spectrum, we calculate analytically the neutrino flux and energies, given the parameter values we have already computed. The all-flavour neutrino flux can be approximately computed as
\begin{equation}
\varepsilon_{\nu, \rm obs} F_{\varepsilon_{\nu, \rm obs}}\approx \frac{3}{8} \Gamma^{4} f_{\rm p \pi}(\varepsilon_{\rm p}) \frac{r_{\rm b}^{2}c \varepsilon_{\rm p}U_{\rm p}(\varepsilon_{\rm p})}{d_L^2} 
\label{eq:neuflux}
\end{equation}
where $\varepsilon_{\nu, \rm obs} \approx \eta_{\rm p \pi}\Gamma \varepsilon_{\rm p}(1+z)$, $\eta_{\rm p \pi}=1/20$ is the fraction of proton energy that is transferred to each neutrino produced, and $U_{\rm p}(\varepsilon_{\rm p})$ is the differential proton energy density at steady state. This can be written as  $U_{\rm p}(\varepsilon_{\rm p})= \varepsilon_{\rm p} n_{\rm p}(\rm \varepsilon_{\rm p})$, where $n_{\rm p}(\varepsilon_{\rm p})=n_{\rm p}(\gamma) /m_{\rm p}c^2$. The steady-state proton distribution $n_{\rm p}(\gamma)$ is given by \citep{Inoue-Tahakara}
\begin{equation}
    n_{\rm p}(\gamma)=Q_{\rm 0}\frac{\gamma_{\rm c}~ t_{\rm p,esc}}{\gamma^{2}} e^{-\frac{\gamma_{\rm c}}{\gamma}} \int_{\rm max[\gamma_{\rm m},\gamma]}^{\gamma_{\rm max}} 
    e^{\frac{\gamma_{\rm c}}{\gamma}}\gamma^{-p} {\rm d}\gamma
    \label{eq:np}
\end{equation}
where $Q_{\rm 0}$ is computed through eq.~\ref{eq:Lp}. The above equation describes essentially a broken power law, i.e., $n_{\rm p}(\gamma)\propto \gamma^{-p}$ for $\gamma_{\rm c}<\gamma<\gamma_{\rm m}$ and $n_{\rm p}(\gamma)\propto \gamma^{-p-1}$ for $\gamma>\gamma_{\rm m}$.

\begin{figure}
\centering
\includegraphics[width=1.1\linewidth]{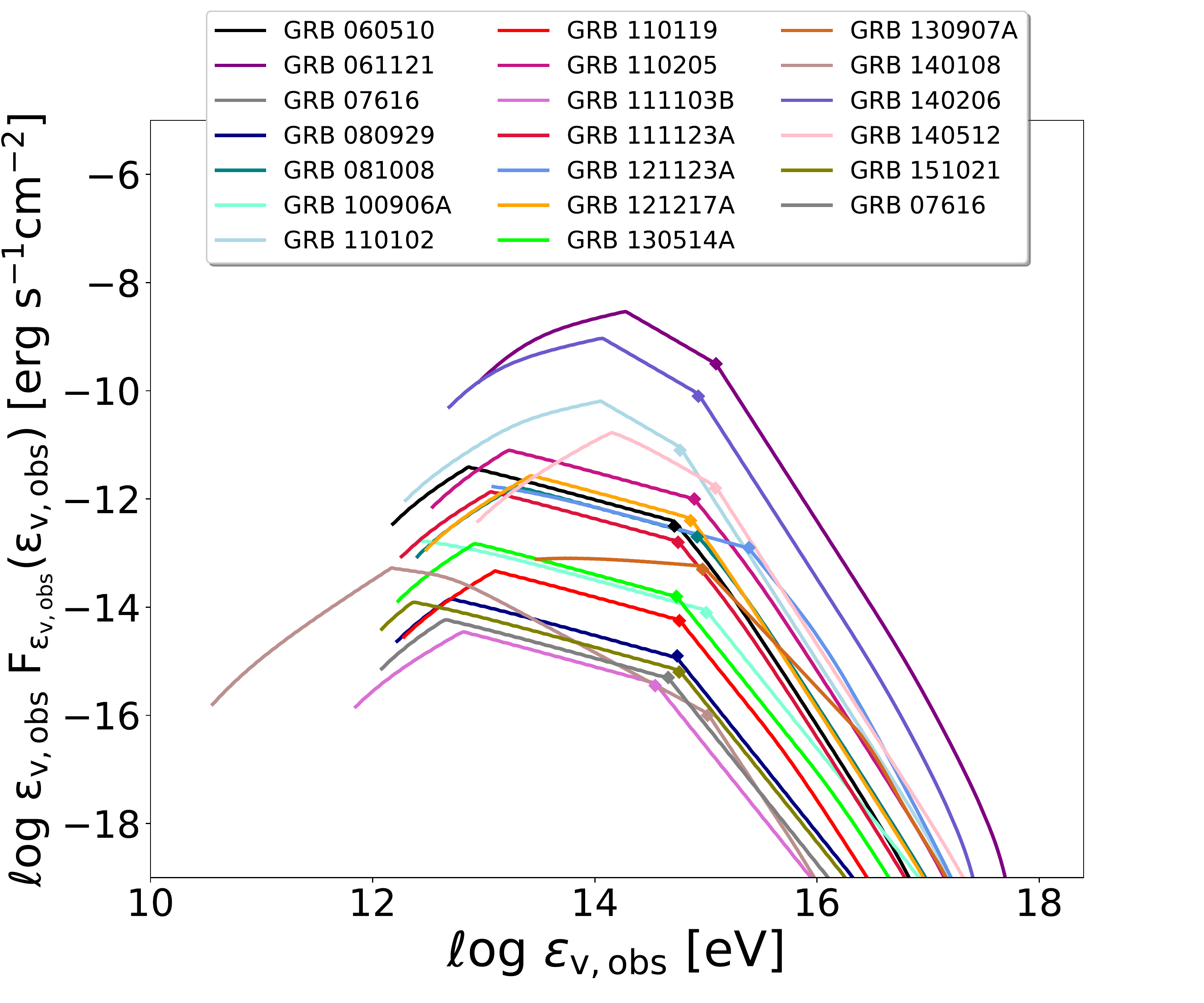}
\caption{Semi-analytical results for the all-flavour neutrino energy flux spectra in the proton synchrotron model for the GRB  sample of \citet{Oganesyan19}, assuming $\Gamma=300$. Results are shown for  the best-fit values of observables from the interval of peak $\gamma$-ray flux. Diamonds mark the characteristic energy above which pions cool due to synchrotron radiation before they decay, leading to a steeper neutrino spectrum. }
\label{fig:fv}
\end{figure}

We apply  eq.~\ref{eq:neuflux} to each GRB of the sample using the parameters that correspond to $\Gamma=300$ (see Table \ref{tab:app} below). The resulting all-flavour neutrino fluxes are shown in Fig.~\ref{fig:fv}. Most of the curves show a triple broken power-law shape that arises from the multiplication of the proton distribution function with the double broken power-law function of photopion efficiency (eq.~\ref{eq.fpg}). The shape of the latter is determined by the proton-synchrotron photon number density (see eq. \ref{eq.nx}).  
A closer look at these results and the observables for each burst shows that 
as the ratio $\gamma_{\rm m}/\gamma_{\rm c}$ decreases and approaches unity, the first branch of the neutrino spectrum disappears.
The neutrino break energy between the first and the second branch of the neutrino energy spectrum occurs typically when protons of Lorenz factor equal to $\gamma_{\rm m}$ interact with photons of energy greater than $E_{\rm pk,obs}$, and corresponds to the peak energy of the neutrino spectrum. This ranges between $10$~TeV and 1 PeV  for $\Gamma=300$, and scales as $\propto \Gamma^{2}$. Because of the strong magnetic fields ($10^6-10^7$~G) involved in the proton-synchrotron model,
the cooling timescale of pions and muons can be faster than their decay timescale (see Appendix \ref{apb}), thus leading to a steepening of the neutrino spectrum \citep[e.g.,][]{waxman-bahcall-97, 2010ApJ...710..346A, Zhang-kumar13}. If $\rm F_{\varepsilon_{\rm v, \rm obs}}\propto \varepsilon_{\rm v, \rm obs}^\chi$, we approximately account for the cooling effects by describing the neutrino flux as $\rm F_{\varepsilon_{\rm v, \rm obs}}\propto \varepsilon_{\rm v, \rm obs}^{\chi-2}$ above an energy that corresponds to that of a pion with equal synchrotron cooling and decay timescales (this is indicated with a diamond in the figure).  While this is a rough approximation~\citep[for a detailed analytical treatment of cooling effects, see e.g.,][]{2012ApJ...752...29H, 2015Tamborra, Pitik2021}, it is sufficient for the purposes of this study. 
More details on the neutrino flux predicted by this model will be presented
elsewhere \citep{Pitik2021}.

\begin{figure}
    \centering
    \includegraphics[width=\linewidth]{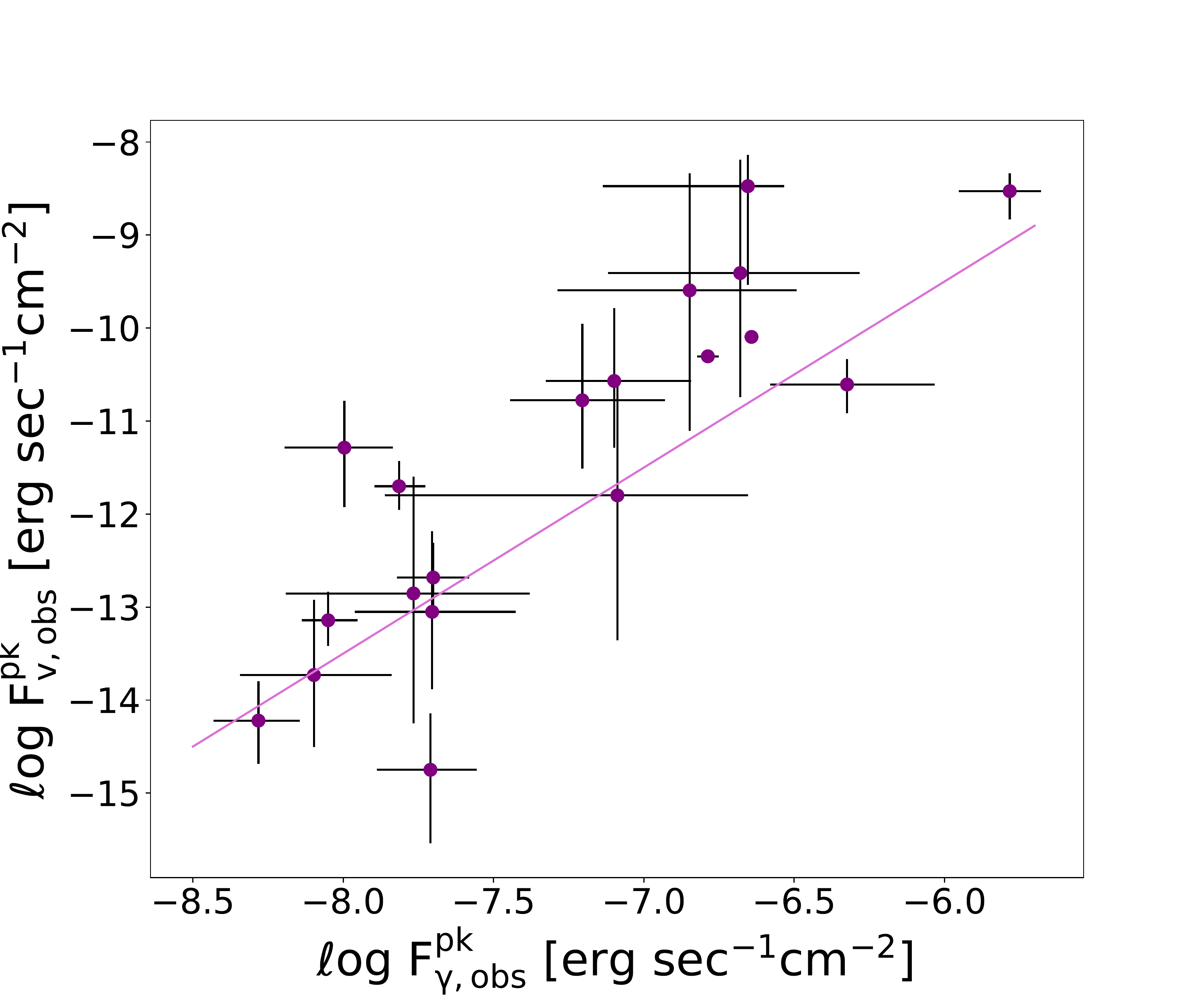}
    \caption{A scatter plot that shows the peak of the predicted all-flavour neutrino energy flux  for the GRBs of the sample, as a function of the ratio of the photon to neutrino peak fluxes, assuming that $\Gamma=300$ (purple circles). The coloured solid line has a slope of two and is plotted in order to guide the eye.}
    \label{fig:scatter}
\end{figure}

We next compare the peak energy fluxes of the all-flavour neutrino and photon spectra, which we denote respectively as $F_{\nu, \rm obs}^{\rm (pk)}$ and $F_{\gamma, \rm obs}^{\rm (pk)}$. Our results are presented in Fig.~\ref{fig:scatter} for $\Gamma=300$. Here, the error bars indicate the 68\% range of values of the respective distributions (see also Fig.~\ref{fig:uncertaintycode}). The predicted peak fluxes of a few GRBs (see \ref{sec:values})  have large uncertainties because at least one of their observables (i.e., $E_{\rm c}, F_{\rm c}, \gamma_{\rm m}/\gamma_{\rm c}$) has large asymmetric error bars. The neutrino peak flux is many orders of magnitude below the peak $\gamma$-ray flux, reflecting the low photopion production efficiency that is expected for conditions that enhance the synchrotron cooling efficiency of protons (see also Fig.~\ref{fig:fboth}). Fig.~\ref{fig:scatter} also shows that for most GRBs the predicted fluxes scale as $F_{\nu, \rm obs}^{\rm (pk)} \propto \left(F_{\gamma, \rm obs}^{\rm (pk)} \right)^2$. Such quadratic relation is expected in the proton-synchrotron model, since both the photopion efficiency and the proton energy density scale linearly with the photon number density (for details, see Appendix~\ref{apdxd}).

\begin{figure*}
    \centering
    \includegraphics[width=0.49\textwidth]{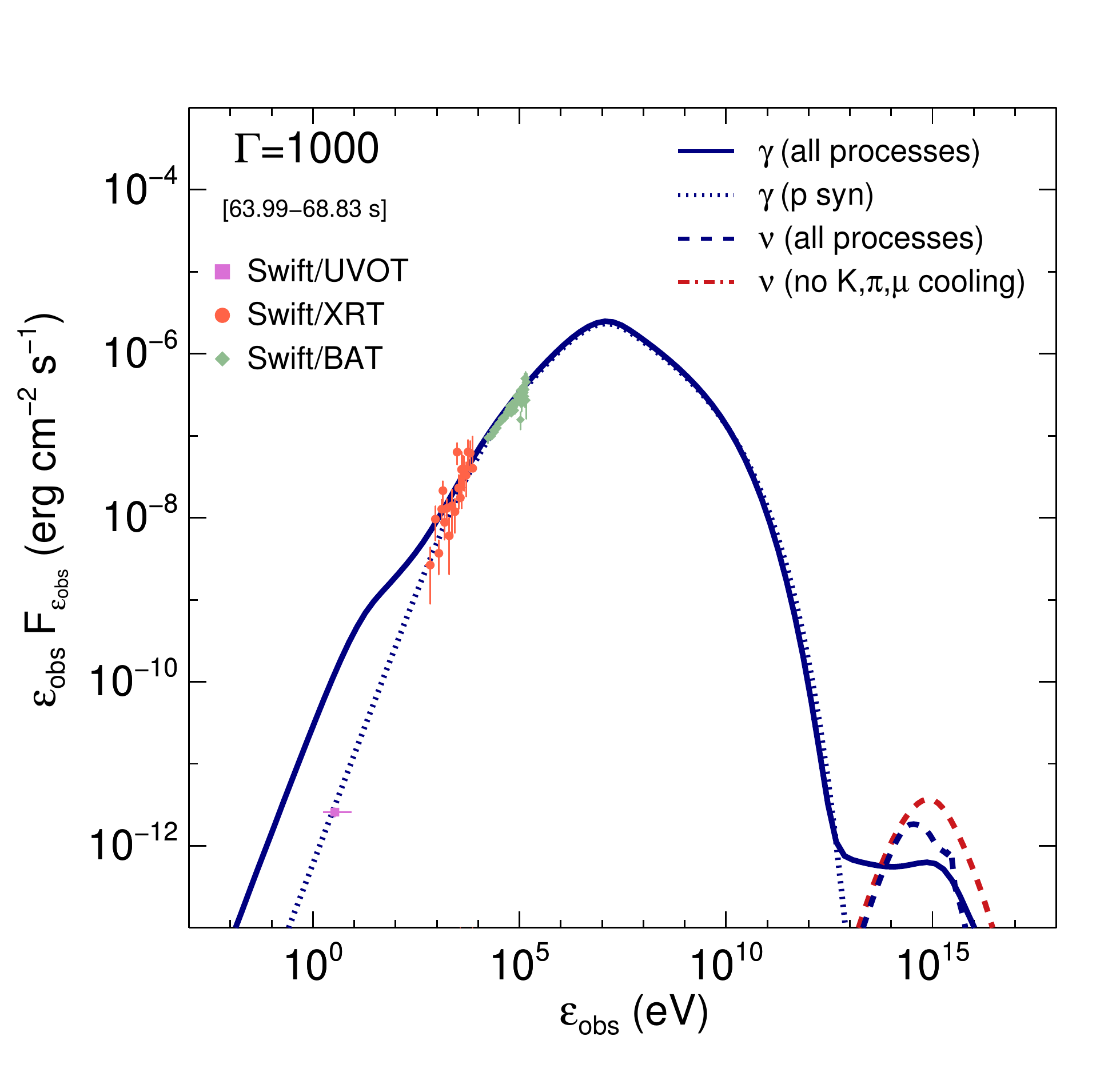}
    \includegraphics[width=0.49\textwidth]{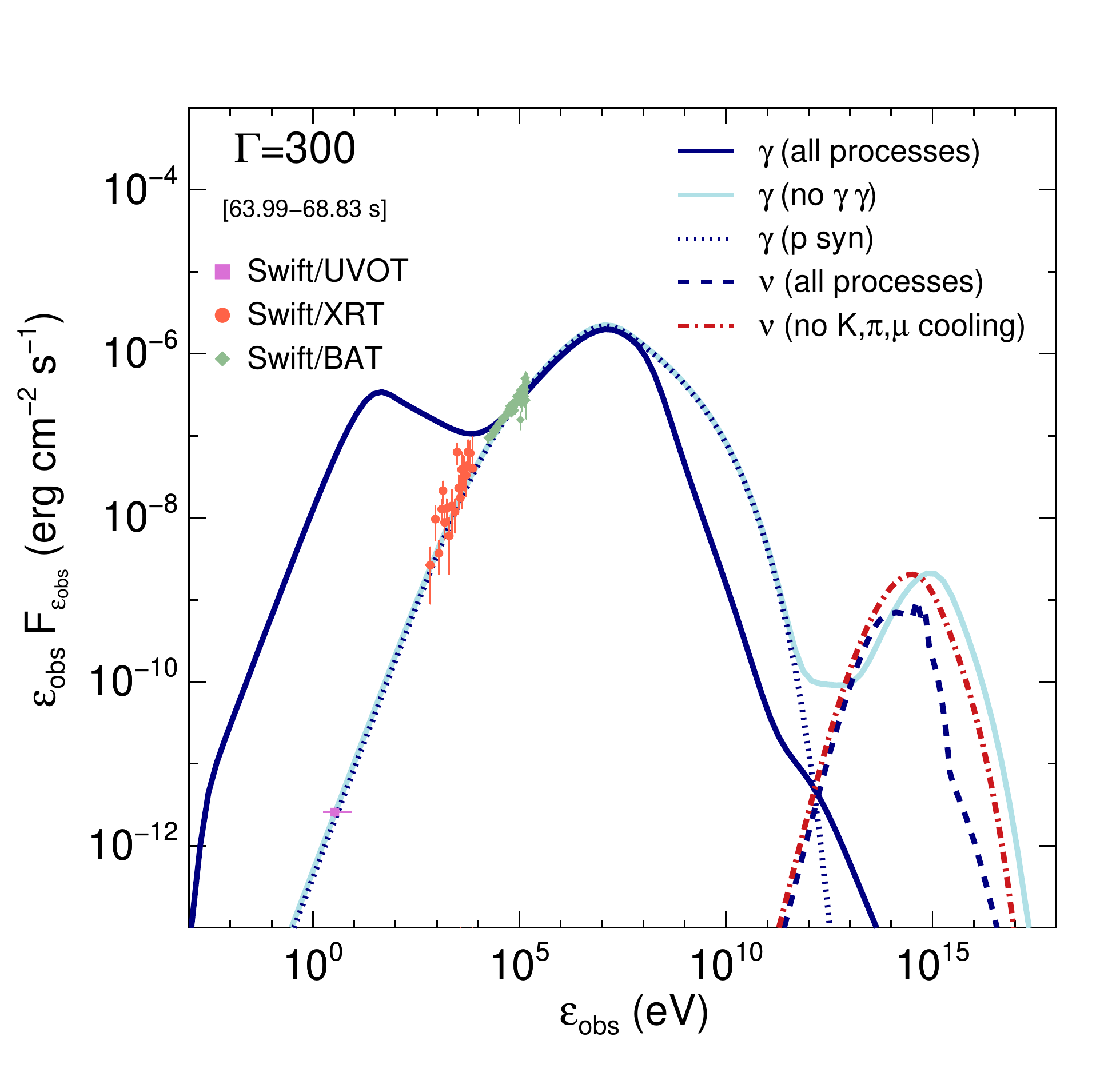}
    \caption{Predicted photon (solid blue line) and all-flavour neutrino (dashed blue line) energy spectra of GRB~061121 in the proton-synchrotron prompt emission model for $\Gamma=1000$ (left panel) and $\Gamma=300$ (right panel). 
    The proton synchrotron spectrum (dotted blue line) and the neutrino spectrum in the absence of meson and muon synchrotron cooling (dashed-dotted red line) are overplotted. In the right panel, the photon spectrum without $\gamma \gamma$ pair production is also shown for comparison (solid light blue line). In both panels, prompt observations by \swift/UVOT, \swift/XRT and \swift/BAT (adopted from \citet{Oganesyan19}) are overplotted with symbols (see inset legend). }
    \label{fig:spec}
\end{figure*}

\subsubsection{Numerical results}
Utilizing the numerical code presented in Sec.~\ref{Sec:3} and the parameter values obtained in Sec.~\ref{Sec:2}, we compute the GRB prompt photon and neutrino spectra in the proton synchrotron scenario  after taking into account all relevant radiative processes. In particular, we demonstrate how the photohadronic interactions, $\gamma \gamma$ pair production, and synchrotron cooling affect the overall photon spectrum of the GRB prompt emission. We also take into account synchrotron cooling of kaons, pions, and muons, as this may be important for certain parameter values (see Appendix \ref{apb}). As an illustrative example we use GRB~061121. Model parameters are inferred by the observables for the first time interval (63.99-68.83~s) analyzed by \cite{Oganesyan19} (see Table B1 and Figure 4 therein), and are summarized in Table~\ref{tab:model} for two indicative values of the Lorentz factor. 

\begin{table}
\centering
\caption{Parameter values used in the numerical proton synchrotron model of GRB~061121.}
\label{tab:model}
\begin{threeparttable}[b]
\begin{tabular}{c ccccc}
\hline
$\Gamma$ & $B$ (G) & $r_{\rm b}$ (cm)  & $\gamma_{\rm m}$  & $L_{\rm p}$ (erg s$^{-1}$) \\ \hline
\hline
300 & $4.9\times10^6$ &  $3.8\times10^{12}$ & $3.2\times10^4$ &  $2.2\times10^{43}$\\ 
1000 & $3.7\times10^6$ & $1.3\times10^{13}$ & $2.1\times10^4$ & $1.9\times10^{41}$\\ 
\hline
\end{tabular}
\begin{tablenotes}
\item Note. -- The parameter values are inferred for the first time interval used in the time-resolved analysis of \citet{Oganesyan19}. Other parameters used are $\gamma_{\max}=10^6$ and $p=2.6$.
\end{tablenotes}
\end{threeparttable}
\end{table}

The broadband photon and all-flavour neutrino spectra are presented in Figure~\ref{fig:spec}. For comparison, we also show the pure proton synchrotron spectrum (dotted lines) that provides a good description of the prompt emission from a few eV to $\sim100$~keV energies. When all processes are taken into account we find a modification of the spectra at $\varepsilon_{\rm obs}< E_{\rm c, obs}\simeq 14$~keV, which becomes stronger for lower $\Gamma$ values. This extra emission can be attributed to ultra-relativistic pairs injected in the source mainly via $\gamma \gamma$ pair production that have cooled down to $\gamma \sim 1$ due to synchrotron radiation in the very strong magnetic field of the emitting region\footnote{In principle, thermalization of electrons due to synchrotron-cyclotron self absorption \citep{Ghisellini_1999} could prevent the electron cooling to $\gamma\sim1$. However, for the parameter values used in this work, the heating rate of electrons is negligible in comparison to their cooling rate.  }. Because of the complete electron cooling down to trans-relativistic energies (i.e., $1.1 \lesssim \gamma \lesssim 2$), cyclotron-synchrotron effects become important at energies $\varepsilon_{*, \rm obs} \lesssim 22~{\rm eV}\, (B/10^6~{\rm G}) (\Gamma/10^3)$. While the numerical code computes correctly the total power lost by a trans-relativistic electron \citep[see e.g., eq.~2 in][]{1998MNRAS.297..348G}, it does not use the appropriate spectrum for the single-particle cyclotron-synchrotron emissivity, which approximately scales as $j_\nu \propto {\rm const}$\footnote{This is still a phenomenological approximation. The true spectrum is more complicated, as it is composed of a series of harmonics \citep[see e.g.,][]{2003A&A...409....9M}.} instead of $\nu^{1/3}$ \citep{ 1997A&A...328...95B,1998MNRAS.297..348G}. As a result, the shape of the low-energy part of the synchrotron spectrum ($\varepsilon_{\rm obs} \lesssim \varepsilon_{*, \rm obs}$) is not described accurately. Because of this limitation of our numerical treatment we cannot exclude the proton synchrotron interpretation for $\Gamma=1000$, where the secondary electron emission is not very luminous and causes deviations from the data below a few eV. 
However, for $\Gamma=300$, the secondary synchrotron emission overshoots the \swift/XRT flux, i.e., at energies above $\varepsilon_{*, \rm obs}\simeq 33$~eV. Given that the main source of secondary pairs comes from the attenuation of proton synchrotron photons above the peak (compare light blue and blue solid lines in the right panel), the secondary emission could not have been avoided even if photohadronic interactions were not taken into account.

In both cases, the neutrino energy spectrum (dashed lines) peaks at PeV energies, but the neutrino peak flux is many orders of magnitude lower than the peak $\gamma$-ray flux (see also Fig.~\ref{fig:scatter}). Because the photopion production efficiency is a sensitive function of the Lorentz factor (see also Fig.~\ref{fig:fboth}), a decrease of $\Gamma$ by a factor of 3 leads to an increase of $\sim2$ orders of magnitude in the neutrino flux. Still, neutrino fluxes comparable to the peak $\gamma$-ray fluxes can be excluded in this scenario because of the strong modification of the prompt emission spectrum by secondaries expected for $\Gamma\lesssim 300$ (with the exact value depending on the properties of individual GRBs). We also show the effects of meson and muon cooling on the all-flavour neutrino spectra (compare dashed blue and dash-dotted red lines). Above the peak energy of the neutrino spectrum, the flux decreases because pions and muons cool before they decay \citep[see also][]{2011Baerwald, 2014MNRAS.442.3026P, 2015Tamborra}.

\begin{figure}
    \centering
    \includegraphics[width=0.49\textwidth, trim=0 20 0 0]{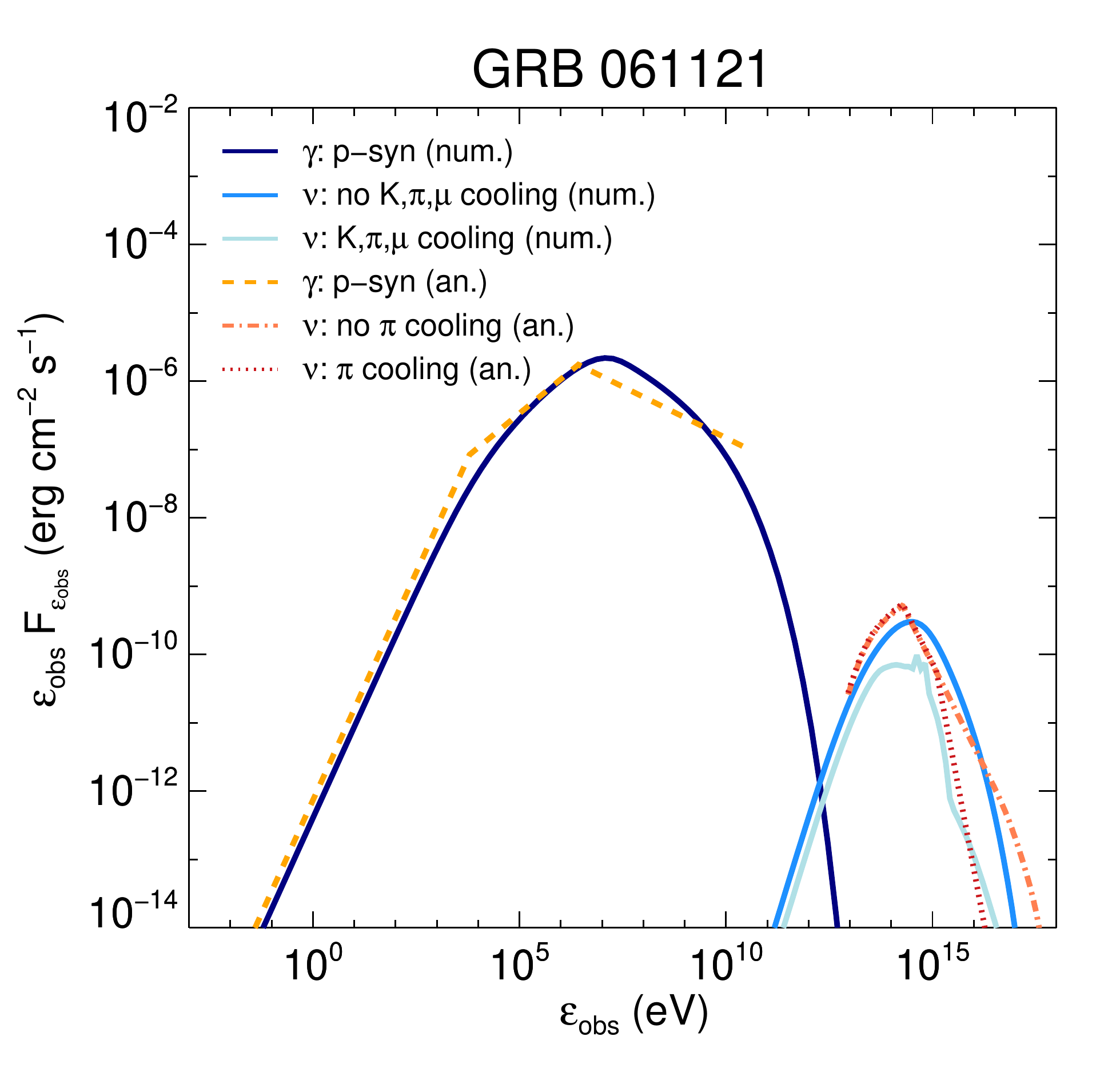}
    \caption{Comparison of numerical and analytical results for the proton-synchrotron spectrum and all-flavour neutrino spectra of GRB~061121 for $\Gamma=300$. For a description of the curves, see inset legend. }
    \label{fig:comparison}
\end{figure}

A comparison of the numerical and semi-analytical results for the neutrino emission is presented in Fig.~\ref{fig:comparison}. For completeness, we also show the respective proton synchrotron spectra, with the analytical one given by eq.~\ref{eq.Fepsilon}. For this illustrative example, we used the parameter values of GRB~061121. In the absence of meson and muon cooling, the peak flux and  peak energy of the numerical and semi-analytical neutrino spectra are in good agreement (compare solid blue and dash-dotted lines). The shape of the numerical neutrino spectrum above the peak energy is smoother than the one found semi-analytically, which is related to the smoothness and curvature of the target photon spectra (compare solid dark blue and dashed lines). Pion cooling leads to a steepening of the neutrino spectrum above a characteristic energy, which is seen in both numerical and analytical spectra (compare solid light blue and dotted lines). However, the peak neutrino flux in the numerical solution is $\sim 4$ times lower than the one of the semi-analytical solution. This drop is not seen in the latter spectrum because  our analytical approach does not take into account the decrease in the number of produced muons (and consequent muon neutrinos) caused by the pion cooling \citep[for a more accurate analytical treatment, see][]{2012ApJ...752...29H}. In this regard, the semi-analytically derived neutrino fluxes (see Sec.~\ref{sec:semianalytic}) are optimistic.

\section{Discussion and Conclusions}
\label{Sec:6}

In this paper we have investigated whether a synchrotron model of relativistic protons could fit observations of prompt GRB spectra (from UV and soft X-rays to MeV $\gamma$-rays). To explain the spectral break seen at a few keV in the prompt spectra of several GRBs \citep{Oganesyan17, Oganesyan19}, we have adopted the marginally fast cooling synchrotron model, according to which the minimum and cooling proton Lorentz factors ($\gamma_{\rm m}$ and $\gamma_{\rm c}$ respectively) are significantly higher that unity and have comparable values. This model requirement may seem like a mere coincidence, since the two characteristic Lorentz factors are set by different physical processes: $\gamma_{\rm m}$ is usually set by the acceleration process, while $\gamma_{\rm c}$ is determined by balancing the cooling timescale with the dynamic timescale of the emitting fluid. Nevertheless, under our working hypothesis, we can infer the magnetic field strength of the emitting region and the minimum energy of the accelerated protons using the best-fit spectral parameters from \cite{Oganesyan19} . 

We have also examined how important are additional physical processes, such as photohadronic interactions and photon-photon pair production, in the overall shape of the GRB spectrum and what is the expected neutrino flux, assuming parameters that lead to proton-synchrotron radiative dominance. We have supported our analytical calculations by utilizing a time-dependent numerical code, which computes the GRB prompt emission spectra taking into account all relevant physical processes without making approximations about cross sections, inelasticities, and production spectra of secondary particles.

Hadronic models have already  been discussed as a viable alternative to the leptonic ones for the high-energy GRB emission, especially after the suggestion that GRBs could be the sources of ultra-high energy cosmic rays (UHECRs) \citep{1995Vietri} and high-energy neutrinos~\citep{Waxman_1997,Murase:2008sp}. One common assumption of these models is that the high-energy part of the spectrum, in the GeV energy band, can be explained by proton synchrotron radiation \citep{Vietri} or by proton-induced cascades \citep{Dermer_2006,Asano_2007}, while the sub-MeV energy band is attributed to synchrotron radiation of accelerated (primary) electrons. If GRBs can accelerate  besides protons heavier nuclei to ultra-high energies, then high-energy $\gamma$-rays can also be produced by ion synchrotron radiation. For source conditions that allow the survival of UHE nuclei in the GRB jet, 
the source is optically thin to photopion production and  very high-energy $\gamma$ rays (TeV photons) are expected to escape the source 
\citep[e.g.,][]{Murase:2008mr,Biehl2018,2020HeinzeBiehl}.


While the aforementioned hadronic models were focused on UHECRs and high-energy $\gamma$-ray emission from GRBs, the goal of our project is to test the hadronic origin of keV-to-MeV GRB emission. Contrary to previous works, the proton spectrum is strongly constrained in the scenario under study. More specifically, the minimum proton Lorentz factor is set according to eq.~(\ref{eq:gm}), motivated by the observational parameters. The proton injection luminosity is also inferred from the observational data, since it is approximated by the observed photon luminosity (see eq.~\ref{eq:Lp}). On the other hand, the maximum proton Lorentz factor is not constrained from the observations, but it is computed by equating the acceleration and synchrotron cooling timescales. Contrary to the studies referred in the previous paragraph, in this project the maximum proton Lorentz factor does not have a significant impact on the model predictions. Lastly, the power-law slope of the proton distribution is much steeper than in other models, as it is related to the photon index of the GRB spectrum above its peak energy (i.e., $p\sim 2.6$ instead of $p \sim 2-2.2$ as usually used). 

A common problem of hadronic models is that protons carry most of the energy, while only a small amount is radiated away. As a result, large proton luminosities are usually required to reproduce the GRB photon emission \citep[e.g.,][]{Asano_2009}. However, 
there is an intrinsic (often overlooked) property of hadronic systems, the so-called hadronic supercriticality, that allows efficient energy transfer from protons to photons. Photohadronic  interactions, namely photopair and photopion production processes, dominate inside the source and essentially drain the  proton energy and transfer it to secondaries, thus increasing the photon efficiency to high values \citep{1992Natur.360..135K, Mast2005, 2012MNRAS.421.2325P}. Different variants of hadronic supercriticalities have been applied to GRBs \citep{2002ApJ...578L..15K, 2014MNRAS.444.2186P, 2018MNRAS.477.2917P}. 
In particular, \cite{2014MNRAS.444.2186P} have shown that the proton synchrotron radiation produces very high-energy $\gamma$-rays that are quenched, producing as a result electron-positron pairs, while the MeV emission is the outcome of Comptonization of photons by cooled pairs. Moreover, a thorough analysis of the parameters of hadronic supercriticality that could be implemented to GRB prompt emission has been discussed in \cite{zoo}. Florou et al 2021 (in preparation) show that hadronic supercriticality can be implemented to GRB prompt emission even in case the source is expanding adiabatically.  All these studies use very different source parameters compared to those of the proton-synchrotron model. Apart from assuming a much smaller minimum proton Lorentz factor ($\gamma_{\rm m}=1$), these works use lower values of magnetic field strength ($B \sim 10^{4}$ G) inside the spherical volume and lower values of proton injection luminosity $L_{\rm p} $ per radius of the emitting volume $r_{\rm b}$, stated as compactness $\ell_{\rm p}$, in order to push the system to the onset of  supercriticality and reproduce a GRB.

In case of a marginally cooled proton synchrotron model the emitting region is compact and strongly magnetised.Similar requirements were found by \cite{2014Beniamini-piran} in the marginally fast cooling electron synchrotron scenario.. This suggests that the dominant form of energy carried by the jet has to be electromagnetic. We therefore compute the Poynting luminosity of a collimated jet, which is defined as:
\begin{equation}
   L_{\rm B, j} =\frac{\Delta \Omega}{4 \pi} c R_\gamma^2 (B\Gamma)^2 \propto  f_{\rm b} \Gamma^{16/3} t_{\gamma, \rm obs}^{2/3} E_{\rm c, obs}^{-2/3}(1+z)^{-4/3}
    \label{eq:LBj}
\end{equation}
where $f_{\rm b}=\theta_j^2/2=5\times10^{-3}(\theta_j/0.1)^2$.
We also compute the jet luminosity due to the relativistic proton population, which is written as:
\begin{equation}
    L_{\rm p, j} = \frac{\Delta \Omega}{4 \pi}  c R_\gamma^2 \Gamma^2 m_{\rm p}c^{2} \int_{\gamma_{\rm c}}^{\gamma_{\rm max}} \gamma n_{\rm p}(\gamma) \rm d\gamma
    \label{eq:Lpj}
\end{equation}
where $n_{\rm p}$ is given by eq.~(\ref{eq:np}). For $\Gamma=300$, the median values for the GRB sample read $L_{\rm B, j}\approx 5 \times 10^{54}$ erg~s$^{-1}$ and $L_{\rm p, j}\approx 5\times 10^{51}$ erg~s$^{-1}$. Given that the Poynting luminosity has only a weak dependence on the observables (see eq.~\ref{eq:LBj}) we can conclude that extreme jet luminosities cannot be avoided in the marginally cooled proton-synchrotron model for GRB prompt emission.

Long-duration GRBs are thought to be associated with the collapse of the core of massive stars \citep{Woosley93,Hjorth+03,Stanek+03,Woosley&Bloom06}. 
The burst is proposed to be powered by the rotation of a rapidly rotating strongly magnetized neutron star (i.e., magnetar) \citep[e.g.,][]{Usov1992,Thompson2004, MetzGian} or by the accretion on to a black hole created after the core collapse \citep[e.g.,][]{Woosley93,macfadyen1999,2001ApJ...550..410M}.  
In the first scenario, the magnetar wind is considered to be the source of the outflow responsible for the relativistic GRB jet \citep{MetzGian}. In the simplest magnetar model, where the jet is solely powered  by the solid-body spin-down energy of the magnetar, the jet luminosity $L_{\rm j}\propto L_{\rm SD}\propto P_0^{-4} B_0^2$, where $P_0$ is  the initial spin period and $B_0$ is the surface magnetic field. For $B_0=10^{15}-10^{16}$~G and $P_0\sim1$~ms, $L_{\rm j}\sim 10^{49}-10^{51}$~erg s$^{-1}$. 
If the central engine is a black hole, accretion in combination to the rotation energy of the black hole can power the GRBs \citep[e.g.,][]{Meszaros1997}. The in-falling stellar material can drag in the large-scale magnetic flux through the progenitor star and the jet can be powered via the Blandford-Znajek  process \citep{BZ1977}. In this scenario, the jet power (equivalent to the power of the central engine) is determined by the magnetic flux through the black-hole horizon $\Phi_{\rm BH}$, namely $L_{\rm j} \equiv L_{\rm BZ} \propto a_{\rm BH} \Phi^2_{\rm BH} M^{-2}_{\rm BH}$, where $a_{\rm BH}$ and  $M_{\rm BH}$ are the BH spin and mass, respectively. As long as the accretion rate $\dot{M}$ is high enough as to sustain the magnetic flux $\Phi_{\rm BH}$ on the BH, the jet power is independent of $\dot{M}$ and approximately constant \citep{2015TchekGiann}. Assuming that the collapsing core of the star forms a black hole with $M_{\rm BH}=4 M_{\odot}$ and $B_{\rm BH}\sim10^{15}$~G at the horizon, the resulting jet power $L_{\rm j}\sim 10^{50}$~erg s$^{-1}$. Based on the jet luminosities expected in collapsars, the jet luminosity predicted by the proton-synchrotron model (see eq.~\ref{eq:LBj}) is rather high.

Apart from the unreasonably high jet luminosity, another problem of the marginally fast-cooling proton synchrotron model for the GRB prompt emission is the modification of the low-energy part of the photon spectrum by secondary pairs and the disagreement with the data, especially for low values of $\Gamma$.
While the proton synchrotron spectrum (see blue dotted lines in Fig.~\ref{fig:spec}) fits perfectly the observations, other physical processes that cannot be neglected (e.g., proton-photon interactions and $\gamma \gamma$ absorption) reshape the photon spectrum in a way that it becomes incompatible with the soft X-ray and optical observations (see blue solid lines in Fig.~\ref{fig:spec}).  Because of the strong inverse dependence of the $\gamma \gamma$ opacity and p$\gamma$ efficiency on $\Gamma$ (see Fig.~\ref{fig:fboth} and Appendix~\ref{addxc}), the contribution of secondaries to the spectrum of GRB~061121 can be suppressed only for $\Gamma \gtrsim 1000$. While the exact value of $\Gamma$ will vary among different GRBs, it will be of the same order. Thus, the cascade emission from secondaries can set very strong lower limits on  the acceptable values of $\Gamma$. These limits may be inconsistent with the bulk Lorentz factors inferred with other methods, e.g., afterglow light curves \citep{2018A&A...609A.112G}, high-energy spectral modeling \citep{2015ApJ...806..194T} and others \citep{Ghirlanda_2011,Nava_2016}.

Our proton synchrotron scenario shows a couple of similarities with the  widely discussed proton-synchrotron models for $\gamma$-ray emission from jetted active galactic nuclei (AGN) \citep{Aharonian2000, 2001APh....15..121M}. Several studies \citep[e.g.,][]{2009sikora,2015Cerruti,2015PD,zharskibotcher15,PVG16,Liodakis_2020} have pointed out the excessive energetic requirements of the proton synchrotron model needed to counterbalance the inefficiency of the proton synchrotron radiation. Moreover, it was shown that photohadronic processes are suppressed, leading to neutrino fluxes from individual AGN that lie well below the IceCube sensitivity threshold \citep[e.g.,][]{2014DimPetMa, 2018ApJ...864...84K, 2019NatAs...3...88G, Liodakis_2020}. While these results are similar to the findings of this work, there are some important differences between the two proton synchrotron models. 
The magnetic field strength in the GRB model considered here is much stronger than the values used in AGN models ($B\sim30-300$~G), as in the latter proton synchrotron cooling is not a requirement. 
This fact also leads to the predominance of the magnetic luminosity over the proton luminosity of the jet in contrast to AGN models where the opposite is noticed. Moreover, the typical energy of protons radiating at the peak of the GRB photon spectrum is several orders of magnitude lower than the one in AGN models. This is related to differences in the comoving magnetic field strengths ($\mathcal{O}(10^6)$~G in GRBs versus $\mathcal{O}(10^2)$~G in AGN), peak energies of the $\gamma$-ray spectra (sub-MeV in GRBs versus GeV in AGN), and bulk Lorentz factors ($\mathcal{O}(10^2)$~G in GRBs versus $\mathcal{O}(10)$~G in AGN). Finally, in the GRB proton synchrotron models the emitted neutrinos are of PeV energies (with the peak determined by the pion and muon synchrotron cooling), while AGN models predict neutrinos in the EeV energy band.

Based on the above discussion, it becomes clear that the
marginally fast-cooling proton synchrotron model for the GRB prompt emission cannot explain the observations presented in \cite{Oganesyan19}. The leptonic synchrotron interpretation, which was put forward by \cite{Oganesyan19}, does not face the problems of the proton-synchrotron model. However, it was disfavoured by \cite{ghisselini20} because the values of magnetic field and radii needed  to reproduce the GRB prompt spectrum with a low Compton ratio suggest a variability timescale much longer than the one observed. Nonetheless, a leptonic synchrotron explanation of the GRB spectra may be viable when studied with a time-dependent approach.
One proposed solution is to include  a spatially decaying magnetic field within an expanding emitting region. The effects of a decaying magnetic  field on the electron and photon spectra have been studied by  \cite{2014NatPh..10..351U}. These authors  showed that the spectral index can become harder than in the case of cooling in a constant magnetic field strength, and the low-energy spectrum below the injection frequency may appear more curved.
Another proposed idea is the assumption of an electro-synchrotron, internal collision induced magnetic reconnection and turbulence (ICMART) model \citep{2011zhang}. In this scenario, the injection rate of electrons   on the synchrotron radiation spectrum increases rapidly with time leading to the   hardening of the corresponding radiation spectrum \citep{Liu_2020}. 
The problems with the marginally fast cooling electron synchrotron interpretation that \cite{ghisselini20} pointed out can be overcome, if the electron
energy losses due to synchrotron are balanced by a continuous source of heating~\citep{2018MBemiamini}. These authors showed 
that there is available parameter space which can produce solutions consistent with the observable constraints, including the variability timescale, if one allows for the possibility of relativistic motion in the jet co-moving frame \citep[e.g., plasmoids from magnetic reconnection][]{2013MNRAS.431..355G, 2016MNRAS.462...48S}. Another idea as a way out is presented by  \cite{Gill_2020}, who investigated the steady-state photon spectra produced by electrons that are either accelerated via magnetic reconnection into a power law or heated via magnetohydrodynamic instabilities in a strongly magnetized relativistic outflow. In their scenario, power-law electrons cool mainly by synchrotron radiation, while the almost monoenergetic electrons from heating cool by Comptonization on thermal peak photons. In both cases, the photon spectra show a low-energy break at $\sim$keV energies that may be consistent with GRB observations \citep{Oganesyan17, Oganesyan19}. In conclusion, the low-energy break seen in the prompt spectra of certain GRBs remains a puzzle.

{\it Note added:} Neutrino predictions of the proton-synchrotron model of GRB prompt emission are also presented in the independent work of \cite{Pitik2021}. Our paper investigates in detail the plausibility of the proton-synchrotron interpretation by analytical and numerical means.

\section*{Acknowledgements}
We would like to thank the anonymous referee for a constructive report, as well as Kohta Murase, Paz Beniamini, and Gabriele Ghisellini for useful comments on the manuscript. We would also like to thank Gor Oganesyan for providing the observational data of GRB~061121 and Georgios Vasilopoulos for discussions on the python  uncertainty routine.
I.F. acknowledges  that  this  research  is  co-financed  by  Greece  and  the  European  Union  (European  Social Fund-ESF) through the Operational Programme ``Human Resources  Development,  Education  and  Lifelong  Learning''  in  the context  of  the  project  ``Strengthening  Human  Resources  Research Potential via Doctorate Research'' (MIS-5000432), implemented  by the  State  Scholarships  Foundation  (IKY).

\section*{Data Availability}
The GRB sample of \citet{Oganesyan19} was used with the best-fit values of observables taken from Table B.1 therein (under the column ``synchrotron model''). Flux points shown in Fig.~\ref{fig:spec} were provided to us by Dr. Oganesyan upon request. Uncertainties have been computed using the python package that is available from \url{https://github.com/kiyami/soad/}. 
Numerical calculations of photon and neutrino spectra were performed making use of the non-public code \texttt{ATHE$\nu$A} \citep{DM12}. Results shown in Figs.~\ref{fig:spec} and \ref{fig:comparison} and any other accompanying outputs of the code are available upon request to the corresponding authors. 


\bibliographystyle{mnras}
\bibliography{bibliography.bib} 




\appendix

\section{Pion \& muon synchrotron cooling}
\label{apb}
The charged byproducts of photohadronic interactions, namely pions and muons,  cannot be always considered to decay instantaneously after their production. If their synchrotron cooling timescale is shorter that their decay timescale, these particles can cool via synchrotron radiation before they decay. They can therefore contribute to the overall GRB photon spectrum, while suppressing the neutrino flux above a certain energy. In this section, we investigate whether pion and muon synchrotron cooling are relevant for the source parameters we use in Sec.~\ref{Sec:4}.

In the case of pions, we calculate the minimum pion Lorentz factor above which the synchrotron cooling timescale is shorter than the decay timescale:

\begin{equation}
\gamma_{\rm \pi}> \sqrt{\frac{6 \pi m_{\rm e}c^{2}}{\sigma_{\rm T} B^{2} \tau_{\rm \pi}} \left(\frac{m_{\rm  \pi}}{m_{\rm e}}\right)^{3}}  
\label{eq:gpi}
\end{equation}
where $\tau_{\rm \pi}=2.6 \times 10^{-8}$ s the charged pion mean lifetime in the lab frame and $m_{\rm \pi}$ its rest mass. The kinematics of this resonance decay predicts a nucleon inelasticity \citep{Mucke_1999}
\begin{equation}
    \kappa=\frac{\gamma_{\rm \pi} m_{\rm \pi} c^{2}}{\gamma m_{\rm p} c^{2}} = 0.2 . \label{eq.a2}
\end{equation}
Therefore the minimum energy of relativistic protons that would produce pions which emit synchrotron radiation before they decay is given by the following relation:
\begin{equation}
\gamma> \gamma^{\rm \pi,syn}_{\rm p} \equiv \left(\frac{m_{\rm  \pi}}{m_{\rm e}}\right)^{5/2} \frac{m_{\rm e}}{m_{\rm p}}  \sqrt{\frac{6 \pi m_{\rm e}c^{2}}{\sigma_{\rm T}  \tau_{\rm \pi}} } B^{-1} \kappa^{-1}
\end{equation}
where the magnetic field $B$ is given by eq. \ref{eq:b}. 

\begin{figure}
    \includegraphics[width=\linewidth]{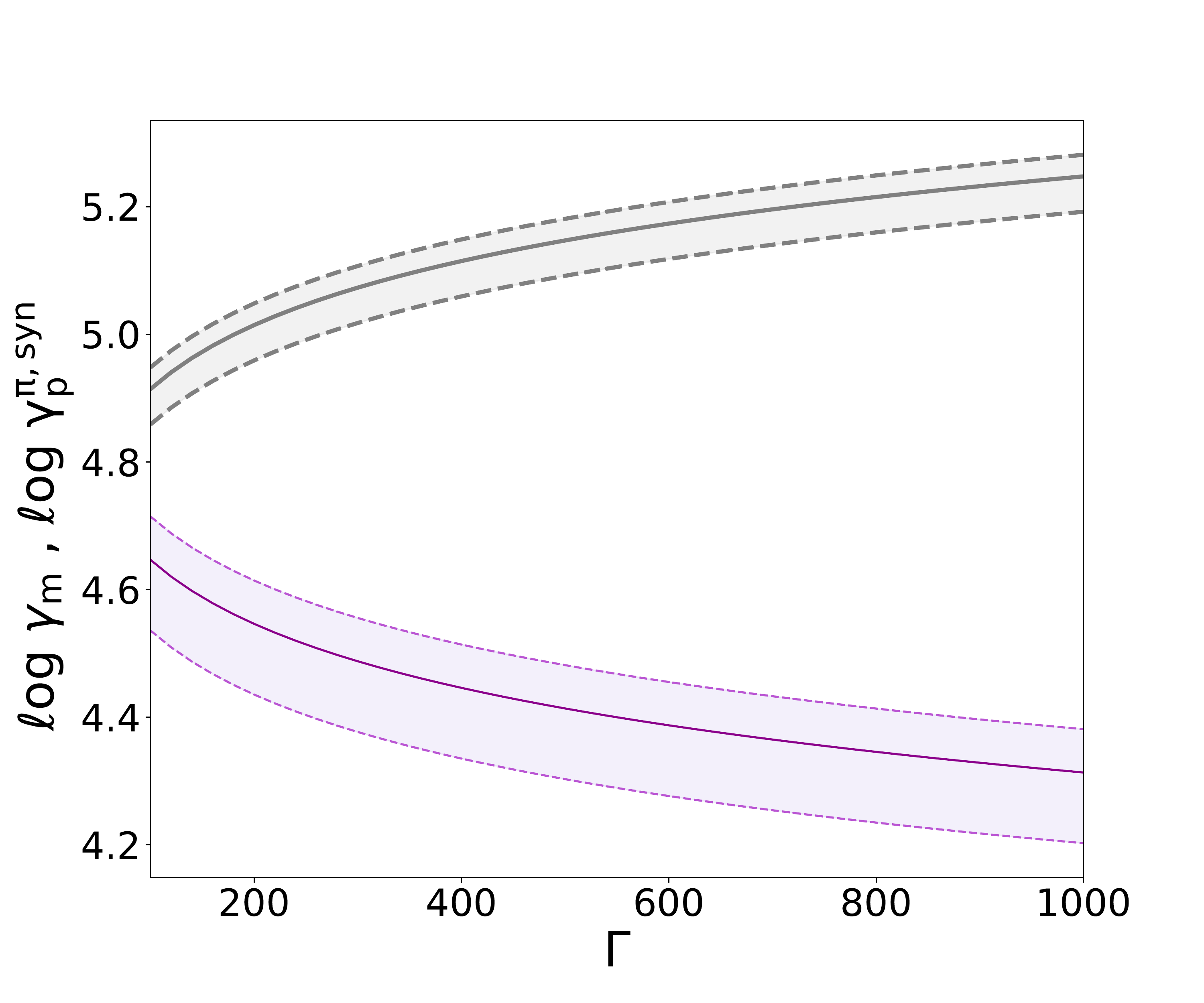}
    \caption{The minimum  Lorentz factor that protons must have in order to produce pions that radiate synchrotron photons before they decay, $\gamma^{\rm \pi, syn}_{\rm p}$,  plotted as a function of $\Gamma$ (solid grey line). Overplotted is also the miminum Lorentz factor of the proton distribution, $\gamma_{\rm m}$ (solid violet line), calculated through eq.~\ref{eq:gm}. Shaded regions indicate the uncertainties on both parameter values. The results are obtained for GRB~061121.}
    \label{fig:gpion}
\end{figure}

In Fig.~\ref{fig:gpion} we plot the minimum values of the proton Lorentz factor required to produce pions that cool via synchrotron before they decay, $\gamma^{\rm \pi,syn}_{\rm p}$, as a function of $\Gamma$ (solid grey line).  In the same figure we also show the minimum Lorentz factor of the proton distribution, $\gamma_{\rm m}$, calculated through eq.~\ref{eq:gm} as a function of $\Gamma$ (solid light blue line). This plot is created using the observables of GRB~061121. The coloured bands indicate the uncertainties on each parameter. These have been calculated following the method of \cite{python-soad}, as described in Sec.~\ref{Sec:2}. We find that the two regions do not intersect for all values of $\Gamma$ considered. Thus, only a fraction of the proton distribution (with $\gamma_{\rm php} \le \gamma \le \gamma_{\max}$) can lead to the production of pions that will cool via synchrotron before they decay. How much pion cooling will affect the observed photon and neutrino spectra will also depend on the shape of the proton energy distribution (e.g., power-law index).  

We follow the same procedure  to compute the minimum proton energy that would produce muons which radiate synchrotron photons before they decay. The muon energy, above which its synchrotron cooling timescale is smaller than its decay timescale, is given by the relation:
\begin{equation}
\gamma_{\rm \mu}>\sqrt{\frac{6 \pi m_{\rm e}c^{2}}{\sigma_{\rm T} B^{2} \tau_{\rm \mu}} \left(\frac{m_{\rm  \mu}}{m_{\rm e}}\right)^{3}} 
\label{eq:gmi}
\end{equation}
where $\tau_{\rm \mu}=2.19 \times 10^{-6}$ s the  muon mean lifetime in the lab frame and $m_{\rm \mu}$ its rest mass. We assume that a muon carries half of the parent pion energy. Therefore the pions that produce such muons have Lorentz factors $\gamma_{\rm \pi} \approx 2 \frac{m_{\pi}}{m_{\mu}} \gamma_{\rm \mu}$. Furthermore, if we take into account eq.~\ref{eq.a2}, we find that the protons that produce such pions have energies:
\begin{equation}
    \gamma>  \gamma^{\mu,\rm syn}_{\rm p} \equiv 1.12~ \sqrt{\frac{6 \pi m_{\rm e}c^{2}}{\sigma_{\rm T}  \tau_{\rm \mu}} \left(\frac{m_{\rm  \mu}}{m_{\rm e}}\right)^{3}} B^{-1}
\end{equation}

\begin{figure}
    \includegraphics[width=\linewidth]{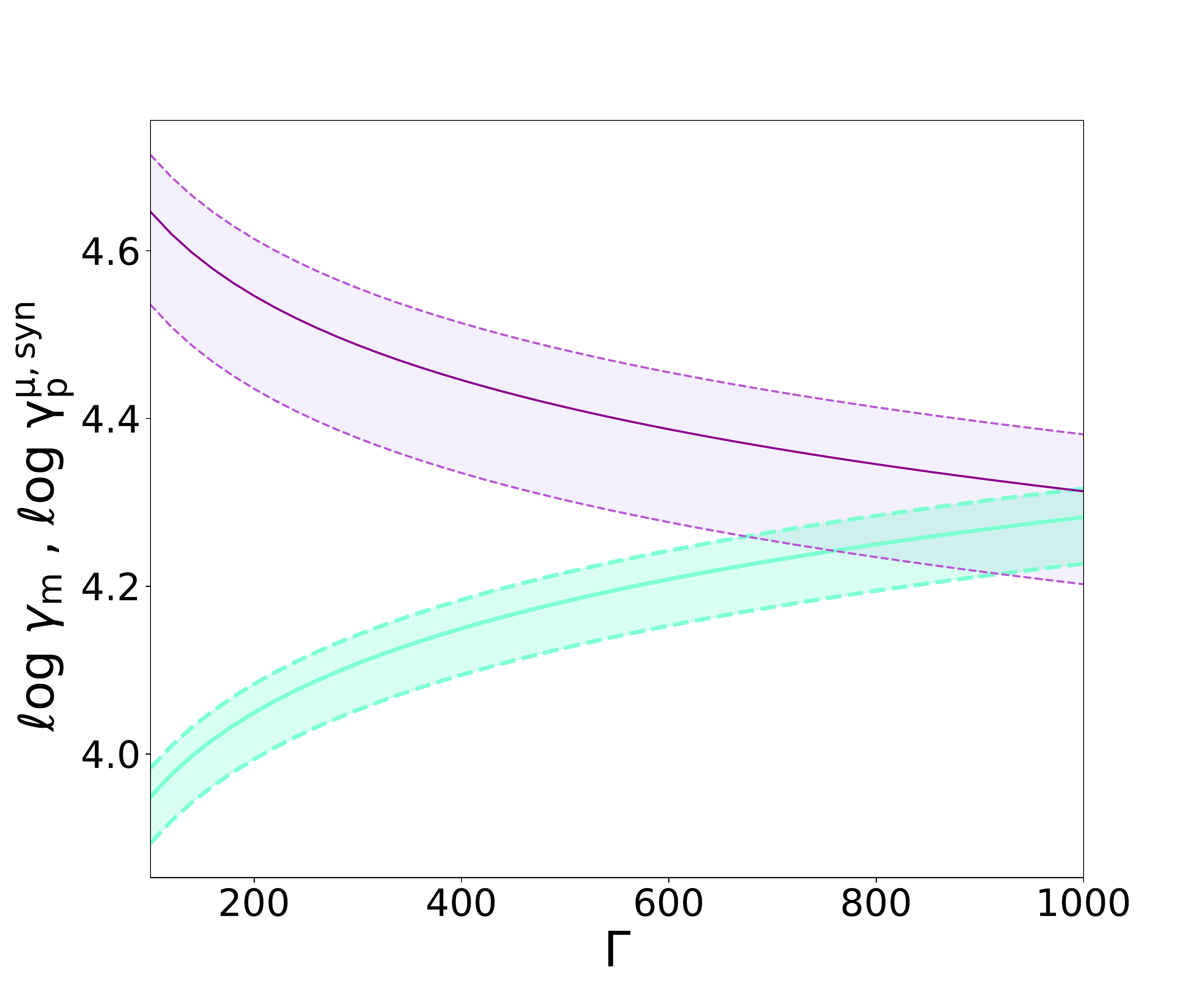}
    \caption{Same as in Fig.~\ref{fig:gpion}, but for the minimum  Lorentz factor that protons must have to produce muons that will cool via synchrotron before they decay (light green shaded region).
    }
    \label{fig:gmuon}
\end{figure}

Figure~\ref{fig:gmuon} displays $\gamma^{\mu,\rm syn}_{\rm p}$ (light  green shaded region) and  $\gamma_{\rm m}$ (purple shaded region) as a function of $\Gamma$ for GRB~061121. The shaded regions have the same meaning as in Fig \ref{fig:gpion}. We conclude that, for all $\Gamma$, protons with $\gamma \gtrsim \gamma_{\rm m}$ may produce muons via photopion interaction that will emit synchrotron radiation before they decay. Thus, muon synchrotron cooling is expected to affect the photon and neutrino spectra.

\section{The $Y$ Compton parameter}
\label{addxc}
Protons can interact with their own synchrotron photons to produce secondary particles via photohadronic processes, as described in the main text. Another 
process that could make protons lose energy while interacting with their own synchrotron photons is inverse Compton (IC) scattering. This is equivalent to the electron synchrotron self-Compton process.

The proton cooling timescale due to IC scatterings in the Thomson regime is written as
\begin{equation}
    t_{\rm p,ic}=\frac{3 m_{\rm e} c^{2}}{4 c \sigma_{\rm T} U_{\rm  ph} \gamma } \left(\frac{m_{\rm p}}{m_{\rm e}}\right)^{3},
    \label{eq.b2}
\end{equation}
where  $\gamma$ is the proton Lorentz factor  and $U_{\rm ph}$ is the 
synchrotron photon energy density. This can be computed through the observed bolometric flux $F_{\gamma}$ (see eq. \ref{eq:Fc}) as \citep{2007Dermer}
\begin{equation}
 U_{\rm ph}= \frac{F_{\gamma} \ 4 \pi r_{\rm b} d_{\rm L}^{2}}{c~\Gamma^{4} \ V_{\rm b}},
\end{equation}
where $ V_{\rm b}$ is the comoving source volume. The proton cooling timescale due to synchrotron and Compton processes can then be written as
\begin{eqnarray}
    t_{\rm p}=\frac{3 m_{\rm e} c^{2}}{4 c \sigma_{\rm T} U_{\rm  B} (1+Y) \gamma} \left(\frac{m_{\rm p}}{m_{\rm e}}\right)^{3} = \frac{t_{\rm p, syn}}{1+Y},
    \label{eq.b3}
\end{eqnarray}
where $U_{\rm  B}=B^{2}/8 \rm  \pi$ is the magnetic energy density, $t_{\rm p, syn}$ is the proton cooling timescale due to synchrotron radiation, and  $Y=U_{\rm ph}/U_{\rm B}$ is the Compton parameter.

\begin{figure}
    \includegraphics[width=\linewidth]{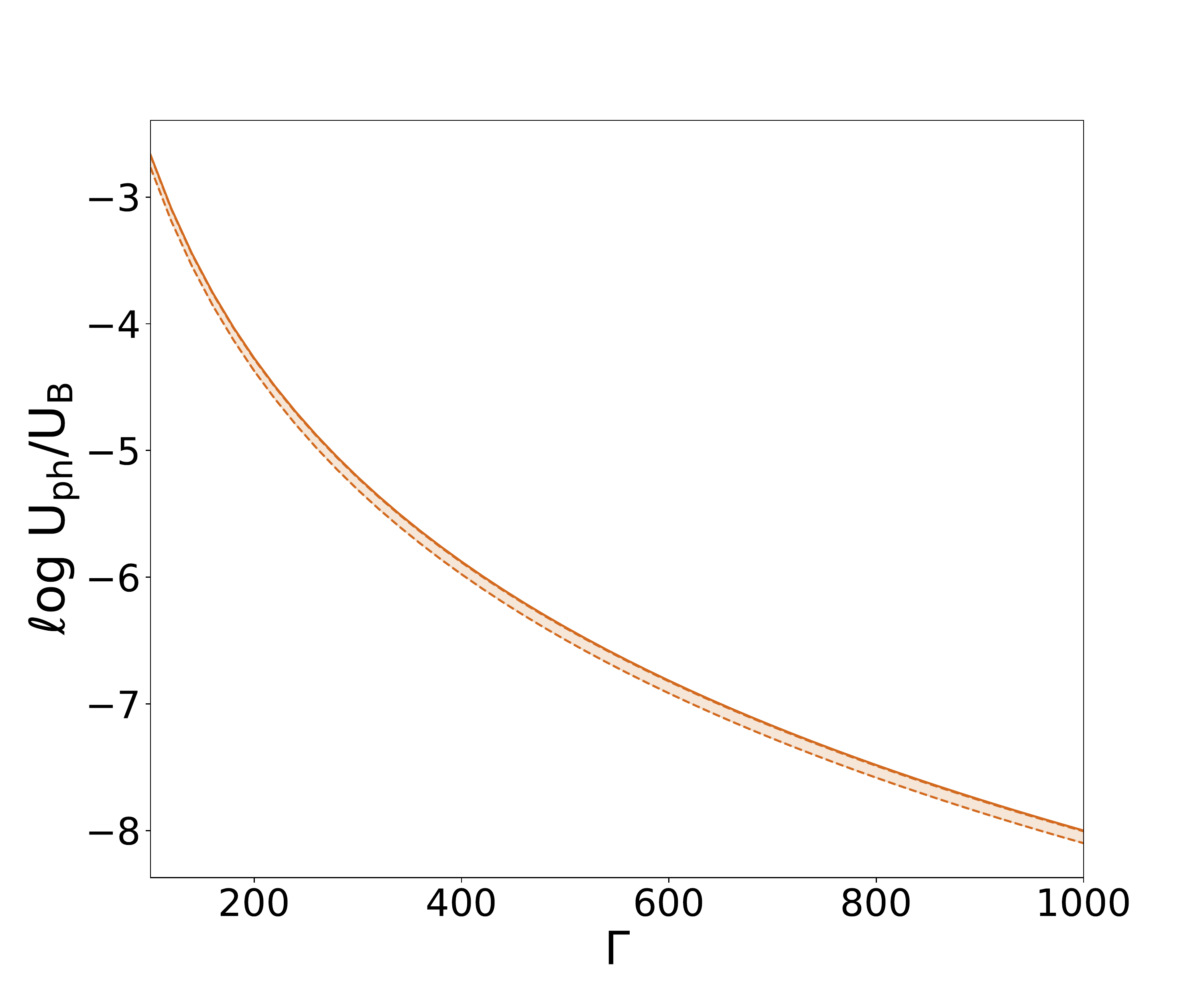}
    \caption{The $Y$ Compton parameter, defined as the energy density ratio of proton synchrotron photons and magnetic fields in the emitting region, as a function of the bulk Lorentz factor.
    Results are shown for GRB~061121. }
    \label{fig:tcool}
\end{figure}

Using the source parameters that we have derived for our GRB sample, namely $B$, $F_{\rm c}$ and $r_{\rm b}$, we compute the photon and magnetic energy densities and the $Y$ parameter as a function of $\Gamma$. An indicative example is shown in Fig.~\ref{fig:tcool} for GRB $061121$. The solid line corresponds to the mean value of $U_{\rm ph}/U_{\rm B}$ while the shaded region indicates the uncertainties. We find that IC scattering on proton synchrotron photons can be safely ignored as a cooling process for the protons in the source for all $\Gamma$ values. Similar results apply to all GRBs in our sample. 

\section{The estimation of photomeson efficiency and $\gamma \gamma$ opacity}
\label{apdxd}

Assuming that the photon distribution in the outflow's rest frame is isotropic the  fractional energy loss rate of a proton with energy $\epsilon_{\rm p}$  due to pion production is written as \citep{Waxman_1997}
\begin{equation}
     t^{-1}_{\rm p\pi}(\gamma)=\frac{c}{2 \gamma^{2}}\int_{0}^{\infty}  {\rm d}x \, x^{-2} n_{\gamma} (x) \int^{2  \gamma x}_{\epsilon_{\rm th}} {\rm d}\epsilon \, \epsilon \sigma_{\rm p\pi}( \epsilon) \xi_{\rm p\pi}(\epsilon),
    \label{eq.fpg}
\end{equation}
where $\epsilon_{\rm th}\simeq 400$ is the threshold energy for production of a $\Delta^{+}(1232)$ resonance,  $\sigma_{\rm p\pi}( \epsilon)\simeq 0.34$~mb for $\epsilon_{\rm th}\le \epsilon \le 980$  is the cross section for pion production for a photon with energy $\varepsilon$ in the proton rest frame (in $m_{\rm e}c^{2}$ units), and $\xi_{\rm p\pi}(\epsilon)\simeq 0.2$ is the average fraction of energy lost by the proton per interaction  \citep{2009herb.book.....D}. Moreover,  $n_{\gamma} (x)$ is the comoving differential photon number density (here, this is the number density of synchrotron photons), and $x=\varepsilon / m_{\rm e} c^{2}$ is the dimensionless photon energy. 

In order to compute the proton energy loss rate for the Bethe Heitler process on the isotropic synchrotron photon field, we follow the equation given by \citep{1970PhRvD...1.1596B}
\begin{equation}
    t_{\rm pe}^{-1}(\gamma)=\frac{3}{8 \pi \gamma} \sigma_{\rm T} c a \frac{m_{\rm e}}{m_{\rm p}} \int_{2}^{\infty} {\rm d}\epsilon  \ n_{\gamma}\left(\frac{\epsilon}{2 \gamma}\right) \frac{\phi(\epsilon)}{\epsilon^{2}} \label{eq.fpe}
\end{equation}
where $a$ is the fine structure constant, $\epsilon = 2 \gamma x$ is the dimensionless photon energy in the proton rest frame also used in eq. \ref{eq.fpg}, and $\phi(\epsilon)$  is a function defined by a double integral, as shown in \cite{1992sikora} (see eqs.~3.13-3.17 therein). 

In both cases, we compute the synchrotron photon number density in the comoving frame 
\begin{eqnarray}
\tilde{n}_{\gamma}(\varepsilon)&=&\frac{3 d_{\rm L}^2  F_{\varepsilon_{\rm obs}}}{\delta^3 r_{\rm b}^{2} c  \varepsilon} \\
n_{\gamma}(x)&=& \frac{3 d_{\rm L}^{2} \ F_{\varepsilon_{\rm obs}}}{\Gamma^3 r_{\rm b}^2 c x}
\label{eq.nx}
\end{eqnarray}
where $\delta \approx \Gamma$, $\varepsilon_{\rm obs} = \Gamma \varepsilon / (1+z)$, $x=\varepsilon/m_{\rm e}c^2$, $\tilde{n}_{\gamma}(\varepsilon)d\varepsilon = n_\gamma(x) d(x)$, and $F_{\varepsilon_{\rm obs}}$ is the  photon flux per unit energy in the observer's frame and is written as
\begin{equation}
F_{\varepsilon_{\rm obs}}=
 \begin{cases}
h^{-1}F_{\rm c} 
\left(\frac{\varepsilon_{\rm obs}}{\rm E_{\rm c,obs}}\right)^{1/3}, \varepsilon_{\rm obs}<E_{\rm c,obs} \\ \\
h^{-1}F_{\rm c} \left(\frac{\varepsilon_{\rm obs}}{\rm E_{\rm c,obs}}\right)^{-1/2}, E_{\rm c,obs}<\varepsilon_{\rm obs}<E_{\rm pk,obs} \\ \\
h^{-1}F_{\rm c} \left (\frac{\varepsilon_{\rm obs}}{\rm E_{\rm pk,obs}}\right)^{-p/2}~ \left(\frac{\rm E_{\rm pk,obs}}{\rm E_{\rm c,obs}}\right)^{-1/2}, \varepsilon_{\rm obs}>E_{\rm pk,obs}.
    \end{cases}
    \label{eq.Fepsilon}
\end{equation}
Here, $F_{\rm c}$ is the flux at the cooling break frequency $\nu_{\rm c, obs}=h^{-1} E_{\rm c, obs}$ (in units of mJy).

The same radiation field that serves as a target for photomeson processes is also a source for $\gamma \gamma$ opacity. The photoabsorption optical depth for a $\gamma$-ray photon of energy $x_{\gamma}$ (in units of $m_{\rm e} c^2$), produced through photopion or photopair processes, in the isotropic synchrotron radiation field $n_{\gamma}(x)$ is:
\begin{equation}
    \tau_{\gamma \gamma} = \frac{r_{\rm b}}{4 \pi}   \int_{0}^{\infty} {\rm d}x ~n_{\gamma}(x)\int  \sigma_{\gamma \gamma} (x,x_{\gamma})~(1-\cos\theta)~ {\rm d}\Omega
    \label{eq.tau}
\end{equation}
where $\sigma_{\gamma \gamma}$ is the pair-production cross section, which is approximated by a step function approximation \citep{1990CoppiBlandford}.
\begin{equation}
   \sigma_{\gamma \gamma} \simeq 0.625 \sigma_{\rm T} \frac{H(x_{\gamma}x(1-\cos\theta)-2)}{x_{\gamma}x (1-\cos\theta)} 
\end{equation}

\bsp	
\label{lastpage}
\end{document}